\documentclass[10pt]{article}
\usepackage{amssymb,amsmath,graphicx}
\usepackage{amsfonts,setspace}
\usepackage{caption}
\usepackage[caption=false]{subfig}
\usepackage{epsfig,latexsym,graphicx}
\usepackage[numbers, sort, compress]{natbib}
\usepackage{setspace}
\begin{document}
	
	\title{\bf Robust Inference for Skewed data in Health Sciences}
	
	\author{Amarnath Nandy, Ayanendranath Basu$^\ast$ and Abhik Ghosh\footnote{Corresponding authors}\\
Interdisciplinary Statistical Research unit\\
		Indian Statistical Institute, Kolkata, India  
		\\
		{\it amarnath\_r@isical.ac.in, ayanbasu@isical.ac.in, abhik.ghosh@isical.ac.in}}
\date{}
\maketitle

\begin{abstract}
Health data are often not symmetric to be adequately modeled through the usual normal distributions; 
most of them exhibit skewed patterns. 
They can indeed be modeled better through the larger family of skew-normal distributions 
covering both skewed and symmetric cases. 
However, the existing likelihood based inference, that is routinely performed in these cases, is extremely non-robust against data contamination/outliers. 
Since outliers are not uncommon in complex real-life experimental datasets, 
a robust methodology automatically taking care of the noises in the data would be of great practical value 
to produce stable and more precise research insights leading to better policy formulation. 
In this paper, we develop a class of robust estimators and testing procedures for the family of skew-normal distributions 
using the minimum density power divergence approach with application to health data. 
In particular, a robust procedure for testing of symmetry is discussed in the presence of outliers. 
Two efficient computational algorithms are discussed. 
Besides deriving the asymptotic and robustness theory for the proposed methods, 
their advantages and utilities  are illustrated through simulations and a couple of real-life applications 
for health data of athletes from Australian Institute of Sports and AIDS clinical trial data.
\end{abstract}

\textbf{Key words:} Skew normal (SN) distribution; Robust minimum density power divergence estimation; 
Wald-type test;  Test for symmetry; Genetic algorithm; Influence function.

\bigskip
\section{Introduction}
\label{SEC:Intro}

Health science is an integral part of medical research where the objective is to improve the quality of human 
(as well as animal) health through appropriate scientific insight generation and necessary policy implementation. 
The backbone of health science research is the efficient analyses of heath data 
obtained from several designed or observational medical experiments or surveys with appropriate target questions in mind. 
The innovations and insights generated from such analyses are essential in medical research to develop cures to any sort of illness
and ensure better health quality; they are also important for any country (and even globally) to prepare appropriate health policies.

\begin{figure}[h]
	\centering
	\includegraphics[width=.6\textwidth]{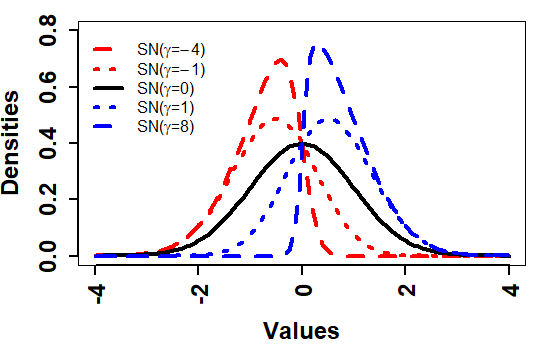}
	\caption{Probability densities of SN distribution with parameters $\mu=0$, $\sigma=1$ 
		and different values of $\gamma$.}
	\label{FIG:SN_density}
\end{figure}

Often, the conventional statistical distribution used in modeling different kinds of health data 
is the bell-shaped and symmetric normal distribution. 
Although it works for some health measurements like height or weight of patients, etc.,  
most health data, specially those measured on some clinical metrics, 
often exhibit empirical skewness so that the conventional normal distribution can not be used to model/analyze them \cite{Partlett:2015}.
Among several possible parametric distributions to model skewed data,
possibly the most popular one is the \textit{Azzalini-type Skew Normal} (SN) distribution family \cite{Azzalini:1985,Azzalini:1986,Azzalini:2005,Azzalini:2011}
which also covers the usual symmetric normal distribution  as a special case; 
see Figure \ref{FIG:SN_density} for a wide variety of distributional shapes (densities) of the SN distribution 
obtained by varying the shape parameter.
Lately, this SN distribution has been successfully applied to model and analyze different types of recent biomedical data
\cite{Castro/etc:2019, Ferreira/etc:2018, Ghalani/Zadkarami:2019, Yiu/Tom:2018, Liu/etc:2016, Smith/etc:2017, Wason/Mander:2015, Gutman/Rubin:2017, Xing/etc:2017, Partlett:2015, Sengupta/etc:2015, Crocetta/Loperfido:2009, vandenHout/Matthews:2009, Bandyopadhyay/etc:2010, Smirnova/etc:2018, Daly/etc:2017, Ngunkeng:2013, Giuntella:2013, Huang/Ku:2010,Hossain/Beyene:2015}.
In this paper, we focus on the SN distribution family for modeling data from different health measurements under one umbrella 
and on the inference  using the estimators of the corresponding SN parameters.

The SN distribution is defined in terms of three parameters, namely the location parameter $\mu \in \mathbb{R}$, 
the scale parameter $\sigma \in \mathbb{R}^{+}$ and the skewness parameter $\gamma \in \mathbb{R}$, and is denoted by $SN(\mu,\sigma,\gamma)$. 
In particular, if $\mu=0$ and $\sigma=1$, it is referred to as the standard SN distribution and is denoted by $SN(\gamma)$. 
The probability density function (pdf) and the cumulative distribution function (cdf) of 
the SN($\mu$,$\sigma$,$\gamma$) distribution are given, respectively, as
\begin{eqnarray}\vspace*{-.5cm}
f_{\boldsymbol{\theta}}(x) &=& \frac{2}{\sigma} \phi\left(\frac{x-\mu}{\sigma}\right)\Phi\left(\gamma\frac{x-\mu}{\sigma}\right),
~~~~ x\in \mathbb{R}, 
\label{EQ:SN_pdf}
\\
F_{\boldsymbol{\theta}}(x)&=&\Phi\left(\frac{x-\mu}{\sigma}\right)-2T\left(\frac{x-\mu}{\sigma},\gamma\right),~~~~ x\in \mathbb{R},
\label{EQ:SN_cdf}
\end{eqnarray}
where $\boldsymbol{\theta}=\big(\mu,\sigma,\gamma\big)^T$ is the vector of unknown parameters, 
$\phi$ and $\Phi$ are the pdf and the cdf of the standard normal distribution, respectively, 
and $T(h,a)$ is Owen's function defined as 
$$
T(h,a)=\frac{1}{2\pi}\int_{0}^{a} \frac{e^{-\frac{h^{2}}{2}(1+x^{2})}}{1+x^{2}}dx,
\hspace{5mm}h,a\in \mathbb{R}.
$$ 
The mean, variance and skewness ($\gamma_{1}$) of a random variable $X$ having SN($\mu$,$\sigma$,$\gamma$) distribution  are  given by
$E(X)=\mu+\sigma\delta\sqrt{\frac{2}{\pi}}$, 
$Var(X)=\sigma^{2}\left(1-\frac{2\delta^{2}}{\pi}\right)$,
with $\delta={\gamma}{(1+\gamma^{2})^{-1/2}}$, and 
$\gamma_{1}=\frac{(4-\pi)\gamma^3}{2\left(\frac{\pi}{2}+(\frac{\pi}{2}-1)\gamma^{2}\right)^{3/2}}.$
Clearly, the SN distribution is positively and negatively skewed according to the sign of the parameter $\gamma$; see Figure \ref{FIG:SN_density}.
At the particular case $\gamma=0$, the SN distribution $SN(\mu, \sigma, 0)$ has skewness zero 
and coincides with symmetric normal distribution, $N(\mu, \sigma^2)$, having mean $\mu$ and variance $\sigma^2$.

Given a random sample $X_1, \ldots, X_n$ from a skewed population, 
we can fit the SN distribution by estimating the parameters $\boldsymbol{\theta}=\big(\mu,\sigma,\gamma\big)^T$ based on the observed data
and the subsequent inference can be done based upon these estimates. 
The usual method of estimation under SN model is the maximum likelihood estimator (MLE) 
which is asymptotically the most efficient at the model.
But, a major drawback of the MLE is its extreme non-robust nature  against data contaminations, outliers or model misspecifications;
this further makes all the MLE based inference highly unstable yielding incorrect insights.  
However, it is not unusual to have some outlying observations in modern complex datasets
due to several external or erroneous factors/activities. 
Hence, a robust inference procedure automatically taking care of the noises (outliers) in the data 
would be of great practical value to produce stable and more precise research insights leading to better policy formulation. 
To further motivate our work in the context of health data analyses, let us consider the following real data example.

\bigskip
\noindent
\textbf{A Motivating Example (AIS data):}\\
We consider the data on health measurements of 706 Australian athletes from 12 different sports 
which were collected at the Australian Institute of Sports (AIS) in 1990
by \citet{Telford/Cunningham:1991} to investigate the relationships of the five routine hematological measures, 
namely the hemoglobin concentration (HC), hematocrit (H), red cell count (RCC), white cell count (WCC) 
and plasma ferritin concentration (PFC) in the blood of these athletes with their height (Ht), weight (Wt) and the sports type.
These measurements are recorded on 1604 occasions from each athletes based on the blood samples collected from their forearm vein 
amidst periods of moderate to intense training but at least 6 hours after a training session. 
Some important derived health measurements like body-mass index (BMI) and lean body mass (LBM) are also reported.
The data were later used by several researchers in different statistical inference problems; 
in particular, few of them  fitted the SN distribution with MLE but only to a few measurements and/or a part of the data 
\cite{Yalcinkaya/etc:2017,Ngunkeng:2013}.

\begin{figure}[h!]
	\centering
	\subfloat[Hemoglobin Conc.~(HC)]{
		\begin{tabular}{c}			
			\includegraphics[width=0.23\textwidth]{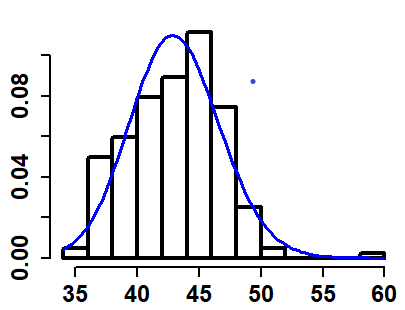}
			\\		\includegraphics[width=0.23\textwidth]{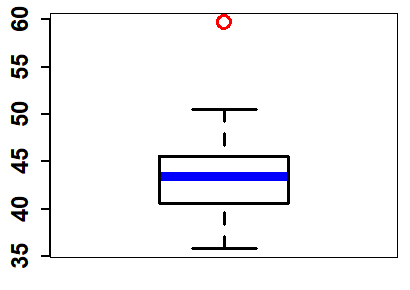}
			\label{FIG:Data_hc}
		\end{tabular}
	}
	\subfloat[Red Cell Count (RCC)]{
		\begin{tabular}{c}			
			\includegraphics[width=0.23\textwidth]{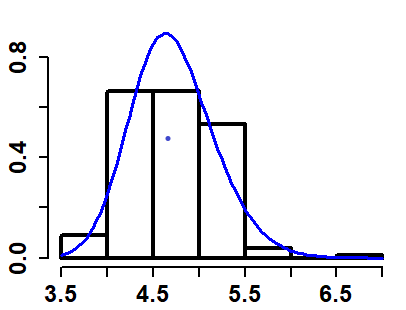}
			\\	\includegraphics[width=0.23\textwidth]{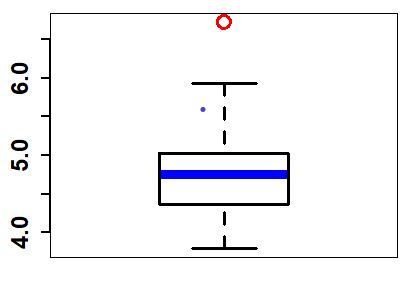}
			\label{FIG:Data_rcc}
		\end{tabular}
	}
	\subfloat[White Cell Count (WCC)]{
		\begin{tabular}{c}			
			\includegraphics[width=0.23\textwidth]{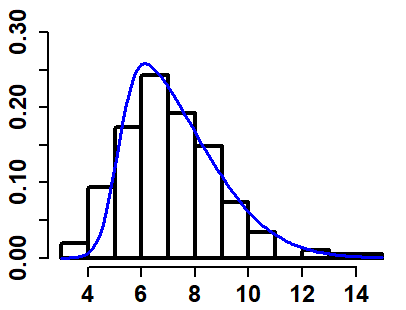}
			\\	\includegraphics[width=0.23\textwidth]{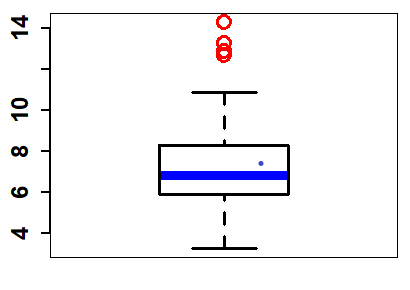}
			\label{FIG:Data_wcc}
		\end{tabular}
	}
	\subfloat[Plasma Ferritin Conc.~(PFC)]{
		\begin{tabular}{c}			
			\includegraphics[width=0.23\textwidth]{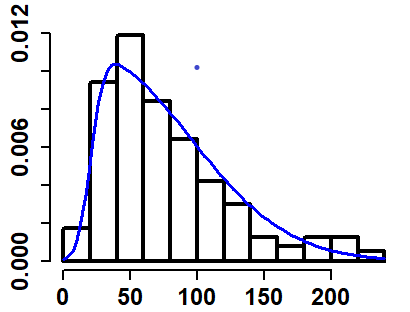}
			\\	\includegraphics[width=0.23\textwidth]{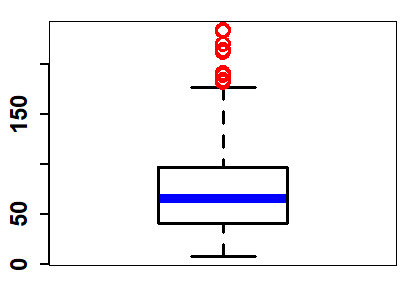}
			\label{FIG:Data_ferr}
		\end{tabular}
	}
	\\
	\subfloat[Body Mass Index (BMI)]{
		\begin{tabular}{c}			
			\includegraphics[width=0.23\textwidth]{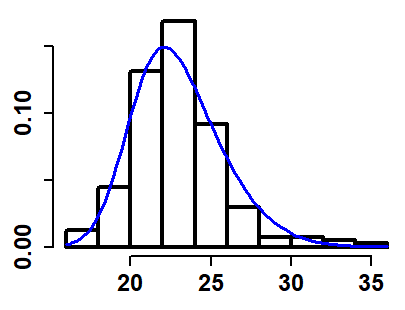}
			\\	\includegraphics[width=0.23\textwidth]{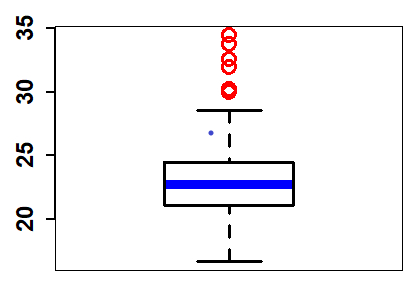}
			\label{FIG:Data_mbi}
		\end{tabular}
	}
	\subfloat[Lean Body Mass (LBM)]{
		\begin{tabular}{c}			
			\includegraphics[width=0.23\textwidth]{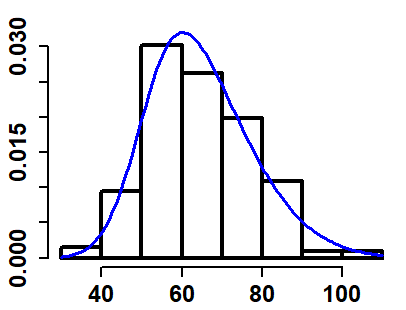}
			\\	\includegraphics[width=0.23\textwidth]{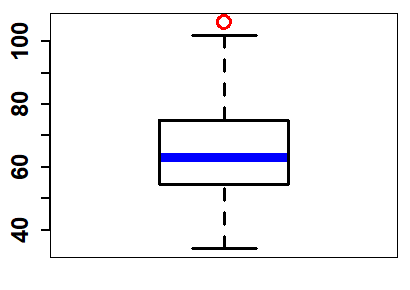}
			\label{FIG:Data_lbm}
		\end{tabular}
	}
	\subfloat[Height (Ht)]{
		\begin{tabular}{c}			
			\includegraphics[width=0.23\textwidth]{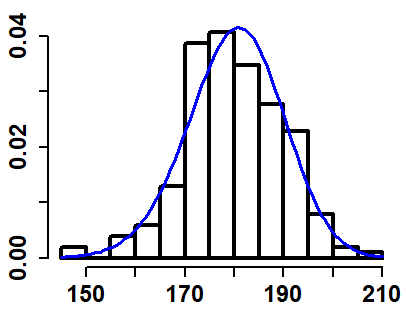}
			\\	\includegraphics[width=0.23\textwidth]{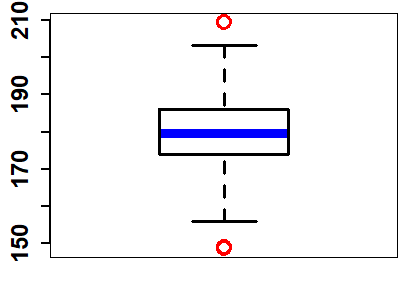}
			\label{FIG:Data_ht}
		\end{tabular}
	}
	\subfloat[Weight (Wt)]{
		\begin{tabular}{c}			
			\includegraphics[width=0.23\textwidth]{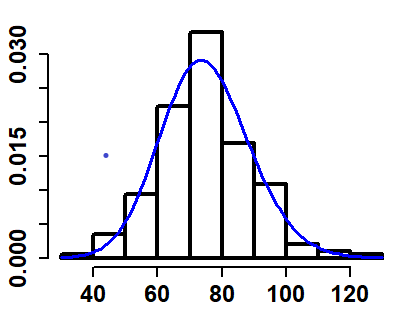}
			\\	\includegraphics[width=0.23\textwidth]{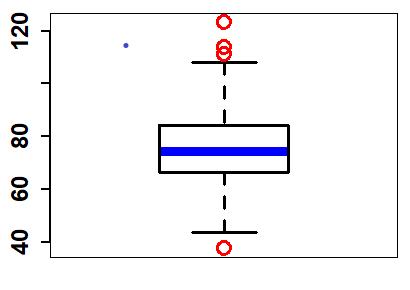}
			\label{FIG:Data_wt}
		\end{tabular}
	}	
	\caption{Histograms and Box-plots of different Health indicator variables from the AIS Data.
		The SN distribution fitted by the MLE is also shown along with the histograms.}
	\label{FIG:Data_hist}
\end{figure}

Let us here consider eight important health measurements, 
namely HC, RCC, WCC, PFC, BMI, LBM, Ht and Wt, from 202 athletes as available in the R package `\textit{DAAG}', 
and plot the corresponding histograms and box-plots in Figure \ref{FIG:Data_hist}.
From the figure it may be seen that several of these variables exhibit clearly skewed patterns. 
For example, RCC, WCC, PFC, BMI and LBM have between mild to prominently pronounced positive skewness. 
Ht, on the other hand, has negative skewness. The other two, HC and Wt are more difficult to judge visually. 
The SN family of distributions, therefore, can be used to model all these health measurements.
However, in all the cases, the respective box-plots reveal one or more outlying values 
which makes the MLE and the associated inference highly unstable. 
The MLE based fits are also shown in the figures along with the histogram, which clearly show the inability of the MLE to 
adequately model the bulk of the data due to the presence of few outlying points. 
In particular, the fitted distributions (by MLE) have a somewhat different mode and skewness 
compared to the majority of the empirical data for the measurements HC, PFC, BMI, LBM and Ht due to strong outlier effects.
We have also computed the MLE after removing the outliers identified through respective box plots 
which are presented in Table \ref{TAB:Data_MDPDE} along with the full data MLE (and their standard errors); 
the changes in the estimates due to the presence of outliers are quite drastic in most cases. 
Although there is only one outlier in HC, its effect is quite dramatic in that 
the deletion of this single observation leads to a reversal in the sign of $\gamma$; 
the MLE of $\gamma$ for the full data (with the outlier) is 0.9655 and any standard testing procedure based on the MLE will reject 
the hypothesis of negative skewness of this distribution although the removal of this single outlier produces a value of $-1.7941$ as the MLE of $\gamma$. 
The MLE of $\sigma$ increases drastically for PFC, BMI and Wt due to the presence of outliers
(73.8403, 4.1327 and 17.6825, respectively, for the full data compared to the corresponding MLEs 57.6705, 2.3489, 
and 13.0243 for the outlier deleted data) and hence results in an inadequate fit for their mode.
These examples clearly illustrate the non-robust and unstable nature of the MLE based inference under the SN distribution 
in the presence of outliers leading to contradictory insights!
\hfill{$\square$}

\bigskip
However, the use of SN distribution for modeling the health measurements and its skewness is indeed justifiable 
through its distributional structure as well as technically from the concept of selective sampling 
\cite{Azzalini/Regoli:2012,Azzalini:2005}. 
For a brief explanation, suppose that we want to model a health measurement variable $U_1$, 
which is assumed to be standardized for simplicity. 
In most cases, health data are collected from a random sample of an appropriately defined subpopulation satisfying a minimum health standard;
in the above example of AIS data all observations are collected from trained athletes who are known to be healthier than others
(in some appropriate health measurement scale). 
Suppose such subpopulation is defined in terms of the condition $U_0>\tau$ for a population random variable $U_0$;
without loss of generality we may assume  $U_0$ to be also standardized. 
Assume $\tau=0$, $U_0$ is normally distributed and has a correlation $\rho$ with $U_1$.
Then, even if $U_1$ is normally distributed in the population, its distribution over the subpopulation, 
i..e, conditional distribution given $U_0>0$ is indeed SN($\gamma$) with $\gamma=\rho(1-\rho^2)^{-1/2}$. 
It can only be symmetric if the target variable is uncorrelated with sub-population defining variable $U_0$.

Therefore the SN distribution is inevitable in health data analyses
and it is the MLE which makes the ultimate inference erroneous under data contamination.
Therefore, it is important to develop an appropriate robust inference methodology for the SN distribution family. 
Unfortunately, little attention has been paid on this issue in the literature, except for a few discrete attempts
for  some particular applications only \cite{Basso/etc:2010, Zeller/etc:2016, Hashimoto:2017,Nurminen/etc:2015}.
In this paper, we develop a simple yet highly efficient robust inference procedure for the SN distribution 
 that can even be generalized to any complex inference problem associated with skewed data quite easily.

Among several approaches to robust inference, we consider the minimum distance approach 
to estimate the parameters of the SN distribution by minimizing an appropriate divergence (distance) measure 
between the data and the model density. 
In particular, we consider the density power divergence (DPD) measure \cite{Basu/etc:1998}
which has lately been extremely popular because of generating extremely robust estimators along with high asymptotic efficiency
\cite{Ghosh/Basu:2013,Ghosh/Basu:2015,Ghosh/Basu:2017,Ghosh/etc:2016,Ghosh/etc:2017,Ghosh:2019,Basu/etc:2011,Basu/etc:2016}. 
In this paper, we first define the minimum DPD estimator (MDPDE) of the parameters of the SN distribution based on a random sample
and discuss its asymptotic properties like consistency and asymptotic normality.
Their asymptotic variance can then be consistently estimated to obtain the standard errors of our proposed estimators
and their robustness properties are discussed through the influence function analysis. 
Since there are complexity in the computation of the MLE itself for SN distribution \cite{Yalcinkaya/etc:2017},
the computation of the MDPDE is also challenging; we have developed an efficient algorithm for this purpose  using 
the concept of Genetic Algorithm \cite{Sivananadam/Deepa:2008}.
We next develop a robust Wald-type test based on the proposed MDPDE along with their asymptotic and robustness properties. 
The important particular case of testing the hypothesis  of symmetry ($\gamma=0$) 
under the SN alternatives are discussed in great detail. The fixed-sample performances of 
the proposed estimation and testing procedures are illustrated through extensive simulation studies. 
Our proposals are then applied to reanalyze the motivating AIS dataset 
as well as to analyze data from an AIDS clinical trial for robust inferential insights.  
Finally the paper ends with some concluding discussion about our work and its possible future extensions.

\section{The Minimum DPD Estimation for the SN distributions}
\label{SEC:MDPDE}

\subsection{Estimating Equation}
\label{SEC:MDPDE_EE}

The DPD family \cite{Basu/etc:1998} is indexed by a single tuning parameter  $\alpha\geq 0$, 
controlling the trade-off between robustness and efficiency.
For two densities $g$ and $f$, both being absolutely continuous with respect to some common dominating measure $\mu$,
the DPD measure between $g$ and $f$ is defined as
\begin{eqnarray}
d_{\alpha}(g, f)&=& \int \bigg\{{f^{1+\alpha}}-\left(1+\frac{1}{\alpha}\right)gf^{\alpha}+\frac{1}{\alpha} g^{1+\alpha}\bigg\} d\mu, 
\hspace{1cm} \mbox{ if }\alpha>0,
\label{EQ:dpd}\\
d_{0}(g, f)&=& \lim\limits_{\alpha\downarrow 0}d_{\alpha}(g, f)
=\int g\log\frac{g}{f} d\mu.
\label{EQ:dpd0}
\end{eqnarray}
Note that, the DPD at $\alpha=0$ is nothing but the well-known Kullback-Leiber divergence (KLD)
associated with the likelihood approach.
We need to minimize the DPD measure between the estimated data density and the postulated model density
to obtain the ``best fitted" model and the corresponding parameter estimates.

Suppose we have a random sample $X_1, \ldots, X_n$ from a population having true density $g$ 
with the associated distribution function $G$ (with the associated measure $\mu$ being the Lebesgue measure). 
We wish to model them by the SN distribution having density $f_{\boldsymbol{\theta}}$ 
and distribution function $F_{\boldsymbol{\theta}}$, which are given in (\ref{EQ:SN_pdf}) and (\ref{EQ:SN_cdf}), respectively.
Then, the minimum DPD estimator (MDPDE) of the unknown model parameter $\boldsymbol{\theta}$ is to be obtained by 
minimizing $d_\alpha(\widehat{g}, f_{\boldsymbol{\theta}})$ over the parameter space $\Theta=\mathbb{R}\times\mathbb{R}^{+}\times\mathbb{R}$,
where $\widehat{g}$ is an estimate of $g$ based on the observed sample. 
One major advantage of the DPD measure is that we can avoid estimating density $g$ by nonparametric smoothing,
which often has several complications like bandwidth selection, curse of dimensionality etc.  
This is because we can rewrite the form of the DPD from (\ref{EQ:dpd}) as 
\begin{eqnarray}
d_{\alpha}(g, f_{\boldsymbol{\theta}})
= \int f_{\boldsymbol{\theta}}^{1+\alpha} d\mu-\left(1+\frac{1}{\alpha}\right) \int f_{\boldsymbol{\theta}}^{\alpha} dG + K,
\nonumber
\end{eqnarray}
where the last term $K=\frac{1}{\alpha}\int g^{1+\alpha}d\mu$ is independent of the parameter $\boldsymbol{\theta}$
and has no effect in our target minimization with respect to $\boldsymbol{\theta}\in\Theta$.
Noting that the second term can be estimated just by plugging in the empirical estimate of $G$, 
namely the empirical CDF obtained based on the sample $X_{1},\ldots,X_{n}$,
the MDPDE can be obtained by minimizing the simpler objective function
\begin{eqnarray}
H_{n}(\boldsymbol{\theta})=\int f_{\boldsymbol{\theta}}^{1+\alpha} d\mu 
- \left(1+\frac{1}{\alpha}\right)\frac{1}{n}\sum_{i=1}^{n} f_{\boldsymbol{\theta}}(X_{i})^{\alpha}.
\nonumber
\end{eqnarray}
For the present case of SN distribution, using the form of $f_{\boldsymbol{\theta}}$ from (\ref{EQ:SN_pdf})
with $\mu$ being the usual Lebesgue measure,
the above MDPDE objective function has the form
\begin{eqnarray}
H_{n}(\boldsymbol{\theta}) 
= H_{n}(\mu,\sigma,\gamma)
&=&\int_{-\infty}^{\infty}\bigg(\frac{2}{\sigma}\bigg)^{\alpha+1}\phi\bigg(\frac{x-\mu}{\sigma}\bigg)^{\alpha+1}\Phi\bigg(\gamma\frac{x-\mu}{\sigma}\bigg)^{\alpha+1}dx
\nonumber\\
&& ~~
-\left(1+\frac{1}{\alpha}\right)\frac{1}{n}\sum_{i=1}^{n}\bigg(\frac{2}{\sigma}\bigg)^{\alpha}\phi\bigg(\frac{x_{i}-\mu}{\sigma}\bigg)^{\alpha}\Phi\bigg(\gamma\frac{x_{i}-\mu}{\sigma}\bigg)^{\alpha}.
\label{EQ:MDPDE_Obj_fn}
\end{eqnarray}
Note that, the integral part of the objective function (\ref{EQ:MDPDE_Obj_fn}) do not have a tractable closed-form expression, 
and hence we need to compute it numerically during the simultaneous minimization of $H_{n}(\mu,\sigma,\gamma)$
with respect to the three parameters $(\mu,\sigma,\gamma)$.
By standard differentiation, we get the estimating equations of our MDPDE as given by
\begin{eqnarray}
\frac{1}{n}\sum_{i=1}^{n} \boldsymbol{u}_{\boldsymbol{\theta}}(X_{i})\bigg(\frac{2}{\sigma}\bigg)^{\alpha}
\phi\bigg(\frac{x_{i}-\mu}{\sigma}\bigg)^{\alpha}\Phi\bigg(\gamma\frac{x_{i}-\mu}{\sigma}\bigg)^{\alpha}
=\boldsymbol{\xi}_\alpha(\boldsymbol{\theta}),
\label{EQ:MDPDE_EstEqn}
\end{eqnarray}
where $\boldsymbol{u}_{\boldsymbol{\theta}}(x)=\frac{\partial}{\partial\boldsymbol{\theta}}\log f_{\boldsymbol{\theta}}(x)$
is the score function of the SN distribution and has the form
\begin{eqnarray}
\boldsymbol{u}_{\boldsymbol{\theta}}(x)
=\Bigg(\frac{x-\mu}{\sigma^{2}}-\frac{\gamma\phi\big(\gamma\frac{x-\mu}{\sigma}\big)}{\sigma\Phi\big(\gamma\frac{x-\mu}{\sigma}\big)},\frac{(x-\mu)^{2}}{\sigma^{3}}-\gamma\frac{(x-\mu)}{\sigma^{2}}\frac{\phi\big(\gamma\frac{x-\mu}{\sigma}\big)}{\Phi\big(\gamma\frac{x-\mu}{\sigma}\big)}-\frac{1}{\sigma},\frac{(x-\mu)}{\sigma}\frac{\phi\big(\gamma\frac{x-\mu}{\sigma}\big)}{\Phi\big(\gamma\frac{x-\mu}{\sigma}\big)}\Bigg)^T,
\label{EQ:SN_score}
\end{eqnarray}
and 
\begin{eqnarray}
\boldsymbol{\xi}_\alpha(\boldsymbol{\theta})
=\int \boldsymbol{u}_{\boldsymbol{\theta}}(x)f_{\boldsymbol{\theta}}(x)^{\alpha+1}dx
=\left(\xi_{\alpha}^{(1)}(\boldsymbol{\theta})~~
\xi_{\alpha}^{(2)}(\boldsymbol{\theta})~~
\xi_{\alpha}^{(3)}(\boldsymbol{\theta})\right)^T.
\label{EQ:xi}
\end{eqnarray}
Clearly there is no closed form solution of the above MDPDE estimating equations in (\ref{EQ:MDPDE_EstEqn})
and we need to solve them numerically in order to obtain the MDPDEs based on a given sample. 
An efficient method for the computation of the MDPDE is dicussed later in Section \ref{SEC:MDPDE_Computation}.

It is important to note that the MDPDE is indeed an M-estimator, since its estimating equation can be written 
in the form $\sum_{i}\psi(X_{i},\boldsymbol{\theta})=\boldsymbol{0}$ for a model based $\psi$-function;
see Equation (\ref{EQ:MDPDE_EstEqn}) to identify it.
Further, as $\alpha\rightarrow 0$, the MDPDE objective function in (\ref{EQ:MDPDE_Obj_fn}) satisfies  
$\left[H_{n}(\boldsymbol{\theta})+\frac{1}{\alpha}\right]\rightarrow 1- $the log-likelihood function, 
and the MDPDE estimating equation in (\ref{EQ:MDPDE_EstEqn}) coincides with the usual score equation
leading to the MLE. Hence, the MDPDEs at $\alpha>0$ can be thought of as a generalization of the MLE 
to achieve greater robustness against data contamination.

\subsection{Asymptotic Efficiency and Standard Error}
\label{SEC:MDPDE_Asymp}

The asymptotic distribution of the MDPDE for the present case of the SN distribution 
can easily be obtained from its general theory or the M-estimation theory. 
In particular, the minimum DPD estimators are $\sqrt{n}$-consistent and asymptotically normal.
At a given $\alpha\geq 0$, if the corresponding MDPDE obtained based on a random sample of size $n$ 
is denoted by $\widehat{\boldsymbol{\theta}}_{\alpha,n}$, and the true parameter value is $\boldsymbol{\theta}_0$, 
we have 
$$
\sqrt{n}\left(\widehat{\boldsymbol{\theta}}_{\alpha,n} - \boldsymbol{\theta}_0\right)
\mathop{\rightarrow}^\mathcal{D} N_3\left(\boldsymbol{0}_3, \boldsymbol{\Sigma}_\alpha(\boldsymbol{\theta})
\right),
$$
where $\boldsymbol{0}_p$ is a $p$-vector with all entries zero and 
$\boldsymbol{\Sigma}_\alpha(\boldsymbol{\theta})= \boldsymbol{J}_\alpha(\boldsymbol{\theta})^{-1}
\boldsymbol{K}_\alpha(\boldsymbol{\theta})\boldsymbol{J}_\alpha(\boldsymbol{\theta})^{-1}.
$
Here, for SN distributions with $\boldsymbol{\theta}=(\mu, \sigma, \gamma)^T$, 
the $3\times 3$ matrices $\boldsymbol{K}_\alpha(\boldsymbol{\theta})$ 
and $\boldsymbol{J}_\alpha(\boldsymbol{\theta})$ are given by
\begin{eqnarray}
\boldsymbol{J}_\alpha(\boldsymbol{\theta})
&=&\int \boldsymbol{u}_{\boldsymbol{\theta}}(x)\boldsymbol{u}_{\boldsymbol{\theta}}^{T}(x)f_{\boldsymbol{\theta}}(x)^{\alpha+1}dx
=\begin{pmatrix}
\begin{array}{ccc}
N_\alpha^{(11)}(\boldsymbol{\theta})  & N_\alpha^{(12)}(\boldsymbol{\theta}) & N_\alpha^{(13)}(\boldsymbol{\theta})\\
N_\alpha^{(12)}(\boldsymbol{\theta})  & N_\alpha^{(22)}(\boldsymbol{\theta}) & N_\alpha^{(23)}(\boldsymbol{\theta})\\
N_\alpha^{(13)}(\boldsymbol{\theta})  & N_\alpha^{(23)}(\boldsymbol{\theta}) & N_\alpha^{(33)}(\boldsymbol{\theta})
\end{array}
\end{pmatrix},
\label{EQ:J}
\\
~~~~~
\boldsymbol{K}_\alpha(\boldsymbol{\theta})
&=&\int \boldsymbol{u}_{\boldsymbol{\theta}}(x)\boldsymbol{u}_{\boldsymbol{\theta}}^{T}(x)f_{\boldsymbol{\theta}}(x)^{2\alpha+1}dx
- \boldsymbol{\xi}_\alpha(\boldsymbol{\theta})\boldsymbol{\xi}_\alpha(\boldsymbol{\theta})^{T}
\nonumber\\
&=&\begin{pmatrix}
\begin{array}{ccc}
N_{2\alpha}^{(11)}(\boldsymbol{\theta}) - \xi_{\alpha}^{(1)}(\boldsymbol{\theta})^2 
& N_{2\alpha}^{(12)}(\boldsymbol{\theta}) - \xi_{\alpha}^{(1)}(\boldsymbol{\theta})\xi_{\alpha}^{(2)}(\boldsymbol{\theta})
& N_{2\alpha}^{(13)}(\boldsymbol{\theta}) - \xi_{\alpha}^{(1)}(\boldsymbol{\theta})\xi_{\alpha}^{(3)}(\boldsymbol{\theta})
\\
N_{2\alpha}^{(12)}(\boldsymbol{\theta}) -\xi_{\alpha}^{(1)}(\boldsymbol{\theta})\xi_{\alpha}^{(2)}(\boldsymbol{\theta})
& N_{2\alpha}^{(22)}(\boldsymbol{\theta}) - \xi_{\alpha}^{(2)}(\boldsymbol{\theta})^2
& N_{2\alpha}^{(23)}(\boldsymbol{\theta}) - \xi_{\alpha}^{(2)}(\boldsymbol{\theta})\xi_{\alpha}^{(3)}(\boldsymbol{\theta})
\\
N_{2\alpha}^{(13)}(\boldsymbol{\theta})  -\xi_{\alpha}^{(1)}(\boldsymbol{\theta})\xi_{\alpha}^{(3)}(\boldsymbol{\theta})
& N_{2\alpha}^{(23)}(\boldsymbol{\theta}) -\xi_{\alpha}^{(2)}(\boldsymbol{\theta})\xi_{\alpha}^{(3)}(\boldsymbol{\theta})
& N_{2\alpha}^{(33)}(\boldsymbol{\theta}) - \xi_{\alpha}^{(3)}(\boldsymbol{\theta})^2
\end{array}
\end{pmatrix},~~~~~~~~
\label{EQ:K}
\end{eqnarray}
where 
$\boldsymbol{\xi}_\alpha(\boldsymbol{\theta})$ and $f_{\boldsymbol{\theta}}$ are as defined in (\ref{EQ:xi})
and (\ref{EQ:SN_pdf}), respectively, 
and 
\begin{eqnarray}
N_\alpha^{(11)}(\boldsymbol{\theta})&=&\int_{-\infty}^{\infty}\Bigg(\bigg(\frac{x-\mu}{\sigma^{2}}\bigg)-\frac{\gamma}{\sigma}\frac{\phi\big(\gamma\frac{x-\mu}{\sigma}\big)}{\Phi\big(\gamma\frac{x-\mu}{\sigma}\big)}\Bigg)^{2}f_{\boldsymbol{\theta}}\big(x\big)^{\alpha+1}dx
\nonumber\\
N_\alpha^{(22)}(\boldsymbol{\theta})&=&\int_{-\infty}^{\infty}\Bigg(\frac{\big(x-\mu\big)^{2}}{\sigma^{3}}-\frac{1}{\sigma}-\frac{\gamma\big(x-\mu\big)}{\sigma^{2}}\frac{\phi\big(\gamma\frac{x-\mu}{\sigma}\big)}{\Phi\big(\gamma\frac{x-\mu}{\sigma}\big)}\Bigg)^{2}f_{\boldsymbol{\theta}}\big(x\big)^{\alpha+1}dx
\nonumber\\
N_\alpha^{(33)}(\boldsymbol{\theta})&=&\int_{-\infty}^{\infty}\Bigg(\frac{x-\mu}{\sigma}\frac{\phi\big(\gamma\frac{x-\mu}{\sigma}\big)}{\Phi\big(\gamma\frac{x-\mu}{\sigma}\big)}\Bigg)^{2}f_{\boldsymbol{\theta}}\big(x\big)^{\alpha+1}dx
\nonumber\\
N_\alpha^{(12)}(\boldsymbol{\theta})&=&\int_{-\infty}^{\infty}\Bigg(\bigg(\frac{x-\mu}{\sigma^{2}}\bigg)-\frac{\gamma}{\sigma}\frac{\phi\big(\gamma\frac{x-\mu}{\sigma}\big)}{\Phi\big(\gamma\frac{x-\mu}{\sigma}\big)}\Bigg)\Bigg(\frac{\big(x-\mu\big)^{2}}{\sigma^{3}}-\frac{1}{\sigma}-\frac{\gamma\big(x-\mu\big)}{\sigma^{2}}\frac{\phi\big(\gamma\frac{x-\mu}{\sigma}\big)}{\Phi\big(\gamma\frac{x-\mu}{\sigma}\big)}\Bigg)f_{\boldsymbol{\theta}}\big(x\big)^{\alpha+1}dx
\nonumber\\
N_\alpha^{(13)}(\boldsymbol{\theta})&=&\int_{-\infty}^{\infty}\Bigg(\bigg(\frac{x-\mu}{\sigma^{2}}\bigg)-\frac{\gamma}{\sigma}\frac{\phi\big(\gamma\frac{x-\mu}{\sigma}\big)}{\Phi\big(\gamma\frac{x-\mu}{\sigma}\big)}\Bigg)\Bigg(\frac{x-\mu}{\sigma}\frac{\phi\big(\gamma\frac{x-\mu}{\sigma}\big)}{\Phi\big(\gamma\frac{x-\mu}{\sigma}\big)}\Bigg)f_{\boldsymbol{\theta}}\big(x\big)^{\alpha+1}dx
\nonumber\\
N_\alpha^{(23)}(\boldsymbol{\theta})&=&\int_{-\infty}^{\infty}\Bigg(\frac{\big(x-\mu\big)^{2}}{\sigma^{3}}-\frac{1}{\sigma}-\frac{\gamma\big(x-\mu\big)}{\sigma^{2}}\frac{\phi\big(\gamma\frac{x-\mu}{\sigma}\big)}{\Phi\big(\gamma\frac{x-\mu}{\sigma}\big)}\Bigg)\Bigg(\frac{x-\mu}{\sigma}\frac{\phi\big(\gamma\frac{x-\mu}{\sigma}\big)}{\Phi\big(\gamma\frac{x-\mu}{\sigma}\big)}\Bigg)f_{\boldsymbol{\theta}}\big(x\big)^{\alpha+1}dx.
\nonumber
\end{eqnarray}
Here we can see that the above integrals do not have a  closed form,
but we can numerically calculate them to compute the asymptotic variance matrix 
at different given values of $\alpha$ and $\boldsymbol{\theta}$. 
Based on these formulas, we can study the asymptotic relative efficiency (ARE) of the proposed MDPDE
which are presented in Table \ref{TAB:ARE} at different values of $\boldsymbol{\theta}=(\mu,\sigma,\gamma)^T$ for the SN distribution.
Note that, these AREs  decrease with increasing $\alpha$
but the loss in efficiency is not quite significant at small positive $\alpha$.

\begin{table}[h]
\caption{Asymptotic Relative Efficiency of the MDPDEs of $(\mu,\sigma,\gamma)^T$ in different SN distributions}
\centering	
	\begin{tabular}{c c|c c c c c c c c}
				\hline
		& & & & & $\alpha$ & & & & \\
		Distribution & & 0(MLE) & 0.05 & 0.1 & 0.2 & 0.3 & 0.5 & 0.7 & 1\\
		\hline
		SN(0,1,1) & $\mu$ & 100 & 99.76 & 98.13 & 94.77 & 86.08 & 77.26 & 68.19 & 58.13\\
		& $\sigma$ & 100 & 99.10 & 95.45 & 91.40 & 86.93 & 76.20 & 64.70 & 52.16\\
		& $\gamma$ & 100 & 98.94 & 95.51 & 92.42 & 90.25 & 84.24 & 76.18 & 65.20\\
		\hline
		SN(0,1,0) & $\mu$ & 100 & 99.41 & 98.22 & 92.81 & 85.34 & 77.00 & 68.92 & 57.23\\
		& $\sigma$ & 100 & 98.87 & 96.11 & 91.09 & 85.66 & 74.39 & 63.91 & 52.82\\
		& $\gamma$ & 100 & 99.24 & 98.57 & 92.86 & 91.34 & 83.76 & 76.82 & 66.95\\
		\hline
		SN(0,1,-1) & $\mu$ & 100 & 99.07 & 96.48 & 91.34 & 85.55 & 79.75 & 68.36 & 58.44\\
		& $\sigma$ & 100 & 98.84 & 95.37 & 90.66 & 84.18 & 76.68 & 65.48 & 54.90\\
		& $\gamma$ & 100 & 98.17 & 94.96 & 91.23 & 89.96 & 81.19 & 72.97 & 65.69\\
		\hline
	\end{tabular}
\label{TAB:ARE}
\end{table}

The above asymptotic variance formula can also help us to obtain the standard errors of the MDPDEs
in any practical application. For the MDPDE $\widehat{\boldsymbol{\theta}}_{\alpha,n}=(\widehat{\mu}_{\alpha,n}, 
\widehat{\sigma}_{\alpha,n}, \widehat{\gamma}_{\alpha,n})^T$, obtained based on a sample of size $n$, 
its standard errors are given by $\sqrt{\Sigma_\alpha^{(11)}(\boldsymbol{\theta}_0)/n}$,  
$\sqrt{\Sigma_\alpha^{(22)}(\boldsymbol{\theta}_0)/n}$ and $\sqrt{\Sigma_\alpha^{(33)}(\boldsymbol{\theta}_0)/n}$, respectively,
where $\Sigma_\alpha^{(ij)}(\boldsymbol{\theta})$ denotes the $(i,j)$-th element of the asymptotic variance matrix 
$\boldsymbol{\Sigma}_\alpha(\boldsymbol{\theta})$ for $i,j=1,2,3$.
A consistent estimate of $\boldsymbol{\Sigma}_\alpha(\boldsymbol{\theta})$
is given by $\boldsymbol{\Sigma}_\alpha(\widehat{\boldsymbol{\theta}}_{\alpha,n})$, 
from which we can easily estimate (consistently) the standard errors of 
the MDPDEs of each parameter $\mu, \sigma$ and $\gamma$.

\subsection{Robustness: Influence Function Analysis}
\label{SEC:MDPDE_IF}

The robustness of an estimator can be theoretically examined through the classical influence function (IF) analysis
\cite{Hampel/etc:1986}. The IF indeed measures the asymptotic (standardized)  bias of the estimator caused by an infinitesimal
contamination at a distant contamination point (say $y$).
Therefore, the boundedness of the IF over the contamination point $y$ restricts the extent of possible bias finitely 
for  the corresponding estimator indicating its robust nature (sometime also referred to as B-robustness to emphasis the boundedness of bias).
On the other hand, an unbounded IF indicates possible unbounded bias and non-robustness of the estimator. 
Further, with similar intuition, the supremum of the absolute IF taken  over all possible contamination points
naturally indicates the extent of (bias) robustness of the corresponding estimator.

\begin{figure}[!b]
	\centering
	\subfloat[IF for $\mu$]{
		\includegraphics[width=0.4\textwidth]{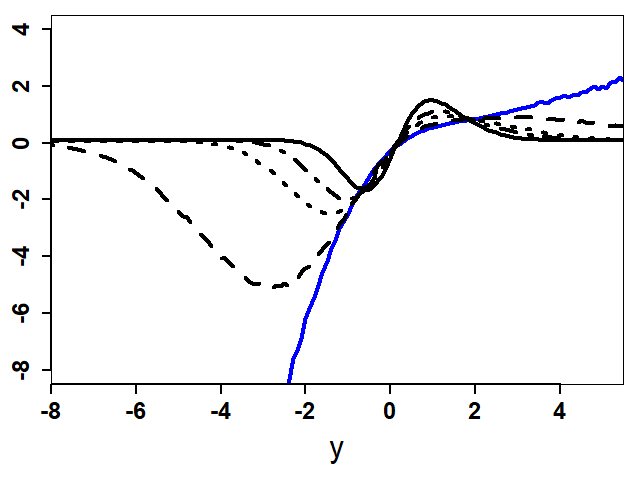}
		\label{FIG:IF_mu}}
	\subfloat[IF for $\sigma$]{
		\includegraphics[width=0.4\textwidth]{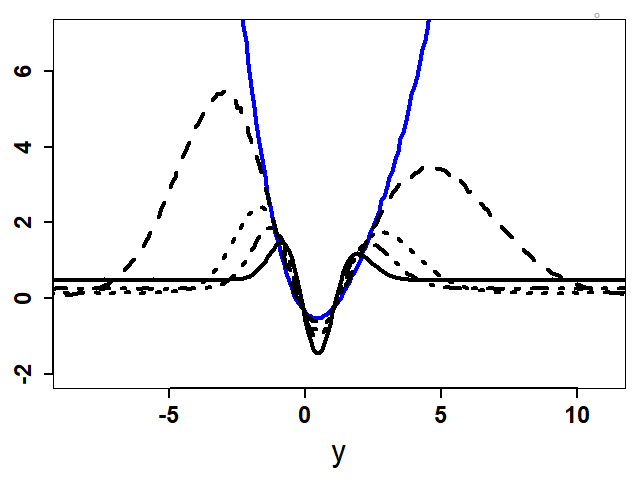}
		\label{FIG:IF_sigma}}\\
	\subfloat[IF for $\gamma$]{
		\includegraphics[width=0.4\textwidth]{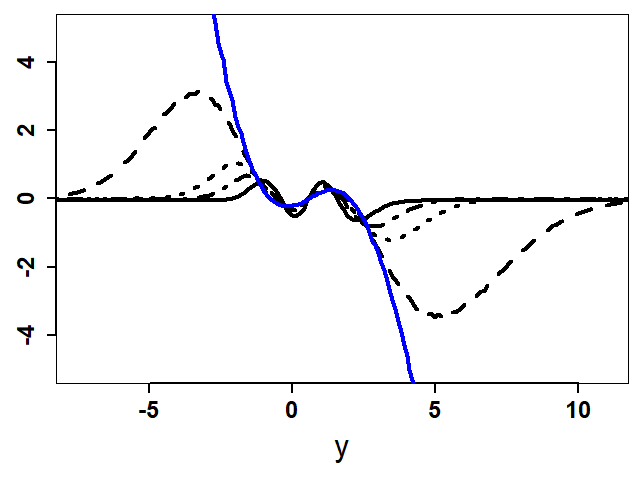}
		\label{FIG:IF_gamma}}
	\caption{Influence functions (IF) of the MDPDEs of the parameters of the SN distribution at SN(0,1,1) for different $\alpha$
		[Blue solid line: $\alpha=0$ (MLE); dashed line: $\alpha=0.1$; dotted line: $\alpha=0.3$; dash-dotted line: $\alpha=0.5$; 
		black solid line: $\alpha=1$]}
	\label{FIG:MDPDE_IF}
\end{figure}

From the theory of M estimator \cite{Hampel/etc:1986} or that of the general MDPDE \cite{Basu/etc:1998,Basu/etc:2011}, 
one can obtain the influence function of the MDPDE functional, say  $\boldsymbol{T}_\alpha$ for a tuning parameter $\alpha$,
under the present case of SN distribution which is given by 
\vspace{-.4cm}
\begin{eqnarray}
IF(y,T_{\alpha},\boldsymbol{\theta})
= \boldsymbol{J}_\alpha(\boldsymbol{\theta})^{-1}
\big[\boldsymbol{u}_{\boldsymbol{\theta}}(y)f_{\boldsymbol{\theta}}(y)^{\alpha} -\boldsymbol{\xi}_\alpha(\boldsymbol{\theta})
\big],
\label{EQ:MDPDE_IF}
\end{eqnarray}
where $\boldsymbol{\xi}_\alpha$ and $\boldsymbol{J}_\alpha$ are defined as in (\ref{EQ:xi}) and (\ref{EQ:J}), respectively.
Now, the form of the SN density $f_{\boldsymbol{\theta}}$ in (\ref{EQ:SN_pdf}) 
and the corresponding score function $\boldsymbol{u}_{\boldsymbol{\theta}}$ in (\ref{EQ:SN_score})
clearly indicates that the above IF is bounded in $y$ for all $\alpha>0$ and unbounded at $\alpha=0$. 
We have presented the IFs of the three parameters $(\mu, \sigma, \gamma)$ in Figure \ref{FIG:MDPDE_IF}
for different values of $\alpha$ which clearly illustrate their boundedness for $\alpha>0$.
This demonstrates the claimed robustness of the MDPDEs at any $\alpha>0$ and the well-known non-robustness of the MLE at $\alpha=0$.
Additionally we can clearly observe the redescending nature of the IFs with increasing values of $\alpha$
which, in turn, indicates the greater extent of robustness with increasing $\alpha$. 

\section{Computation of the MDPDE}
\label{SEC:MDPDE_Computation}

In order to compute the MDPDE, we need to minimize the objective function in (\ref{EQ:MDPDE_Obj_fn})
simultaneously with respect to the three parameters $(\mu, \sigma, \gamma)$,
or equivalently solve the three estimating equations given in (\ref{EQ:MDPDE_EstEqn}).
These are not straightforward numerical exercises due to the complex form of the objective function
and standard numerical procedures like Newton-Raphson algorithm fail.
It is indeed also a known problem in case of the  computation  of the MLE for the SN distribution as well,
for which some advanced numerical procedures has been tried in the literature.
Here, we describe two possible efficient algorithms for the computation of our MDPDE at any given $\alpha>0$.

\subsection{Genetic Algorithm}
\label{SEC:GA}

The genetic algorithm (GA) has been successfully applied for the computation of the MLE 
under the SN model by \cite{Yalcinkaya/etc:2017}.
The GA is an useful and appropriately designed  randomized search technique 
to find exact or approximate solutions in an optimization problem.
Although John Holland first introduced this algorithm in 1960,
it has become popular lately through the works of David Goldberg and others with divergent applications 
\cite{Goldberg:1980,Sivananadam/Deepa:2008}.
The name unsurprisingly came from its structural similarity with genetic mutations and crossover across generations
following the basic principle of the Darwinian Theory of ``Survival of the Fittest".
For an optimization problem, we need to consider an appropriate fitness function 
(often the objective function itself) which produce the fitness value of each possible (candidate) solution under the objective criterion. 
Then, in brief, the GA starts with an initial set of candidate solutions (chromosomes)
and iterates over the subsequent generations to produce new sets of solutions (chromosomes) through recombination and mutation
where the solutions with better fitness values have a higher chance to be there in the subsequent generation
so that the objective function is improved towards optimality. 

To compute the MDPDE using GA, we consider the objective function $H_{n}(\boldsymbol{\theta}) = H_{n}(\mu,\sigma,\gamma)$
as the fitness function with lower values indicating greater fitness of the solution vector $H_{n}(\boldsymbol{\theta} = (\mu,\sigma,\gamma)$.
Then, the algorithm traverses through the following steps.

\bigskip\noindent
\textbf{\underline{GA for Computation of the MDPDE}:}
\begin{itemize}
	\item[\textbf{Step 1.}] We start with an initial set of $N$ candidate solutions denoted as 
	$\mathcal{P}^{(0)} =\{ \boldsymbol{\theta}_{1}^{(0)},\boldsymbol{\theta}_{2}^{(0)},...,\boldsymbol{\theta}_{N}^{(0)}\}$.
	Set $m=0$.
	
	\item[\textbf{Step 2.}] Compute the fitness function $H_{n}(\boldsymbol{\theta})$ for each solutions in 
	$\mathcal{P}^{(m)} =\{ \boldsymbol{\theta}_{1}^{(m)},\boldsymbol{\theta}_{2}^{(m)},...,\boldsymbol{\theta}_{N}^{(m)}\}$.
	
	\item[\textbf{Step 3.}] From the set $\mathcal{P}^{(m)}$, we choose some parent solutions to generate new solutions (offsprings)
	through the `\textit{Fitness Proportionate Selection}' scheme, where the probability of selection 
	is proportional to the (better) fitness values.\\ 
	Alternative schemes like `\textit{Roulette Wheel Selection}' or `\textit{Tournament Selection}' \cite{Chudasama/etc:2011} can also be used. 
	
	\item[\textbf{Step 4.}] We form a new set of $N$ candidate solutions, denoted as  
	$\mathcal{P}^{(m+1)} =\{ \boldsymbol{\theta}_{1}^{(m+1)},\boldsymbol{\theta}_{2}^{(m+1)},...,\boldsymbol{\theta}_{N}^{(m+1)}\}$,
	for the next iteration (generation) using the following two steps:
	
	\begin{itemize}
		\item[\textbf{Step 4.1.}] We choose a specific number (say $N_E$) of elite solutions (survivor) from $\mathcal{P}^{(m)}$
		which are carried forward over the next iteration (generation) without any alterations. 
		They are again chosen by the criterion of having best fitness  values.
		
		\item[\textbf{Step 4.2.}] For generating remaining $(N-N_E)$ candidate solutions for next generations, 
		we perform crossover and mutation operations (through some weighted combination) to the solutions from $\mathcal{P}^{(m)}$ 
		according to some pre-specified crossover probability ($P_C$) and mutation probability ($P_M$).
		The crossover leads the solutions to a convergence while mutation increases diversity among the solutions
		to avoid being stuck at a local optima. 
	\end{itemize}

	\item[\textbf{Step 5.}] Set $m=m+1$ and go to \textbf{Step 2}.
	
	\item[\textbf{Step 6.}] Repeat \textbf{Step 2} to \textbf{Step 5}, until an appropriate (pre-specified) convergence criteria is satisfied.\\
	When stopped, the fittest solution in the last iteration (generation) is returned as the optimal solution (the MDPDE). 
\end{itemize}

Note that, in order to implement the above G, we need to first specify the necessary tuning parameters $N, N_E, P_C, M_P$;
it is suggestive to take $N_E$ as $5\%$ of $N$, a higher value of $P_C$ and a lower values of $P_M$ for faster convergence \cite{Sivananadam/Deepa:2008}.
In all our numerical experiments (simulation studies), 
we have used the R package `\textit{GA}' to implement the Genetic Algorithm
with $N=50$, $N_E=2$, $P_M=0.1$, $P_C=0.8$ and a maximum of 5000 iterations (generations) as stopping criterion. 
However, one challenge using this approach is to choose appropriate values of these tuning parameters
for any real life application!

\subsection{Gradient Descent Method}
\label{SEC:GD}

The method of gradient descent is another popular first-order iterative optimization  algorithm 
mostly used in Machine Learning \cite{Kim/etc:2016,Snyman/etc:2018}. 
To find the minimum of the an objective function, this method progresses iteratively by updating the parameter values 
taking steps proportional to the negative of the gradient (first order derivative) of the objective function. 
For choosing these steps in each iteration, there are various types of algorithms available in the literature 
\cite{Vandenberghe/etc:2019, Robinns/etc:1951}. It is important to note here that this gradient descent approach
might converge to just to a local minimum depending on the initial parameter value considered; 
however, if the function is convex, which is mostly the case for our MDPDE, 
we expect to achieve the global minimum by starting with any reasonable initial value. 

Considering again the MDPDE objective function $H_{n}(\boldsymbol{\theta}) = H_{n}(\mu,\sigma,\gamma)$,
the gradient descent algorithm can be used to find its minimum, i.e., the required MDPDE, through the following steps:

\bigskip\noindent
\textbf{\underline{Gradient Descent for Computation of the MDPDE}:}
\begin{itemize}
	\item[\textbf{Step 1.}] Start with an initial parameter value $\theta_{0}$=($\mu_{0}$,$\sigma_{0}$,$\gamma_{0}$)
	and a step size (tuning parameter) $\lambda>0$.\\	Set $m=0$. 
	
	\item[\textbf{Step 2.}] Calculate $\nabla H_{n}(\boldsymbol{\theta_{m}})$, 
	the derivative of the function $H_{n}(\boldsymbol{\theta})$ with respect to $\boldsymbol{\theta}$
	evaluated at the point $\boldsymbol{\theta_{m}}$ (the solution at the $m^{th}$ step of iteration).
	
	\item[\textbf{Step 3.}] Update the solution at $(m+1)^{th}$ step as: 
	$\boldsymbol{\theta}_{m+1}=\boldsymbol{\theta}_{m}-\lambda \nabla H_{n}(\boldsymbol{\theta_{m}}).$

	\item[\textbf{Step 4.}] Set $m=m+1$ and go to \textbf{Step 2}.

	\item[\textbf{Step 5.}] Repeat \textbf{Step 2} to \textbf{Step 4}, 
	until an appropriate convergence criteria is satisfied.
\end{itemize}

Here, we only need to choose one tuning parameter $\lambda$ for the gradient descent algorithm 
and there exist several suggestions for its optimum selection; see, e.e, \cite{Barzilai/etc:1988, Yuan/etc:1999}. 
For all our numerical illustrations here, we have taken $\lambda$=0.04,
and the initial parameter value to be the maximum partial likelihood estimates of $\boldsymbol{\theta}=(\mu,\sigma,\gamma)$, 
obtained by using the R function `\textit{sn.mple}', and the convergence criterion as no significant 
(relative) change in the objective function. It has been observed through the extensive simulation studies
that both the gradient descent and genetic algorithm perform quite similarly for the computation of the MDPDE
under SN model, with the gradient descent taking significantly less computation time. 
Accordingly, the real data applications are performed using gradient descent algorithm only.

\section{Robust Wald-type tests based on MDPDE}
\label{SEC:Test}

\subsection{General Theory for Composite Hypotheses}
\label{SEC:Test_Gen}

We now consider the problem of testing statistical hypotheses. 
Suppose that, based on random sample $X_1, \ldots, X_n$ from the SN distribution,
we want to test the composite hypothesis
\begin{eqnarray}
H_{0} : \boldsymbol{\theta} \in \Theta_{0} \hspace{2mm} \mbox{against} \hspace{2mm} 
H_{1}: \boldsymbol{\theta} \notin \Theta_{0},
\label{EQ:Hyp_Comp}
\end{eqnarray}
for some closed subset $\Theta_0$ of the parameter space $\Theta$. 
In most applications, the restricted (null) parameter space $\Theta_{0}$ is defined by a set of $r\leq 3$ restrictions of the form 
$\boldsymbol{m}\big(\boldsymbol{\theta}\big)=\boldsymbol{0}_{r}$,
where $\boldsymbol{m}:\Theta \mapsto \mathbb{R}^{r}$ is a known function. 
We assume that the $3\times r$ matrix 
$\boldsymbol{M}(\boldsymbol{\theta})= \frac{\partial \boldsymbol{m}^{T}(\boldsymbol{\theta})}{\partial \boldsymbol{\theta}}$ 
exists, is continuous in $\boldsymbol{\theta}$ and rank$[M(\boldsymbol{\theta})]=r$.
The simplest possible case is $\Theta_{0}=\{\boldsymbol{\theta}_0\}$ for some fixed $\boldsymbol{\theta}_0=(\mu_0, \sigma_0, \gamma_0)\in\Theta$, 
where $r=3$, $\boldsymbol{m}(\boldsymbol{\theta})=\boldsymbol{\theta}-\boldsymbol{\theta}_0$ and $\boldsymbol{M}(\boldsymbol{\theta})=\boldsymbol{I}_3$,
the identity matrix of order 3. Other common cases for the SN distribution could be testing for one or two parameters (among three)
considering the remaining parameter(s) as nuisance. 
As noted earlier the usual test based on the MLE is non-robust 
and hence we discuss an Wald-type test based on the robust MDPDEs following \citet{Basu/etc:2016}.

If $\widehat{\boldsymbol{\theta}}_{\alpha,n}$ denote the MDPDE of $\boldsymbol{\theta}$ based on the given sample,
the Wald-type test statistic for testing the hypothesis (\ref{EQ:Hyp_Comp}) is given by 
\begin{eqnarray}
W_{\alpha,n}=n \boldsymbol{m}^{T}\big(\widehat{\boldsymbol{\theta}}_{\alpha,n}\big)
\big[\boldsymbol{M}^{T}\big(\widehat{\boldsymbol{\theta}}_{\alpha,n}\big)\boldsymbol{\Sigma}_\alpha\big(\widehat{\boldsymbol{\theta}}_{\alpha,n}\big)
\boldsymbol{M}\big(\widehat{\boldsymbol{\theta}}_{\alpha,n}\big)\big]^{-1} \boldsymbol{m}\big(\widehat{\boldsymbol{\theta}}_{\alpha,n}\big), 
\label{EQ:TestStat_gen}
\end{eqnarray}
where $\boldsymbol{J}_{\alpha}$ and $\boldsymbol{K}_{\alpha}$ are as defined in (\ref{EQ:J}) and (\ref{EQ:K}), respectively. 
At $\alpha=0$, this Wald-type test statistics coincides with the usual Wald test based on the MLE.

From the asymptotic distribution of the MDPDE $\widehat{\boldsymbol{\theta}}_{\alpha,n}$ in Section \ref{SEC:MDPDE_Asymp},
it immediately follows that $W_{\alpha,n}$ asymptotically follows a (central) chi-squared distribution, $\chi_{r}^{2}$,
with $r$ degrees of freedom under the null hypothesis in (\ref{EQ:Hyp_Comp}). 
Therefore, we reject $H_{0}$ in (\ref{EQ:Hyp_Comp}) at $\tau_0$ level of significance if 
$W_{\alpha,n}\geq \chi_{r,\tau_0}^{2}$,  the upper $(1-\tau_0)$-th quantile of $\chi_r^2$ distribution.

From the general theory of \citet{Basu/etc:2016}, the MDPDE based Wald-type test is consistent at any fixed alternatives.
Under the contiguous hypothesis of the form $H_{1,n}:~\boldsymbol{\theta}_{n}=\boldsymbol{\theta}_{0}+n^{-1/2}\boldsymbol{d}$, 
with $\boldsymbol{\theta}_0\in\Theta_{0}$ and $\boldsymbol{d}\in\mathbb{R}^3\setminus\{\boldsymbol{0}_3\}$,
$W_{\alpha,n}$ asymptotically follows a non-central chi-squared distribution, denoted as $\chi_{r,\delta}^{2}$, 
having $r$ degrees of freedom and the non-centrality parameter
$\delta = \boldsymbol{d}^{T}\boldsymbol{Q}_\alpha\big(\boldsymbol{\theta}_{0}\big)
\boldsymbol{d}$,
with $\boldsymbol{Q}_\alpha\big(\boldsymbol{\theta}_{0}\big) = \boldsymbol{M}\big(\boldsymbol{\theta}_{0}\big)
\big[\boldsymbol{M}^{T}\big(\boldsymbol{\theta}_{0}\big) \boldsymbol{\Sigma}_{\alpha}\big(\boldsymbol{\theta}_{0}\big)
\boldsymbol{M}\big(\boldsymbol{\theta}_{0}\big)\big]^{-1} \boldsymbol{M}^{T}\big(\boldsymbol{\theta}_{0}\big)$.
Based on this result, an approximate expression of contiguous power function of the test based on $W_{\alpha,n}$ 
can be calculated as 
$
\Pi_{\alpha}\big(\boldsymbol{\theta}_{n}\big)=1-G_{\chi_{r,\delta}^{2}}\big(\chi_{r,\tau_0}^{2}\big),
$
where $G_{\chi_{r,\delta}^{2}}$ is the cdf of the $\chi_{r,\delta}^{2}$ distribution.

The robustness properties of the MDPDE based Wald-type tests were first discussed by \cite{Ghosh/etc:2016} for general parametric models,
which also hold for our SN distribution case. For completeness, we restate the main results briefly.
At the null distribution with $\boldsymbol{\theta}_0\in\Theta_0$, 
the first order IF of the Wald-type test statistic is inconclusive (identically zero)
but the second order IF has the form
\begin{eqnarray}
IF_{2}\big(y,W_{\alpha,n},{\boldsymbol{\theta}_{0}}\big)
=2 IF\big(y,T_{\alpha},{\boldsymbol{\theta}_{0}}\big)^{T}\boldsymbol{Q}_\alpha\big(\boldsymbol{\theta}_{0}\big)
IF\big(y,T_{\alpha},{\boldsymbol{\theta}_{0}}\big).
\label{EQ:IF2_test_Gen}
\end{eqnarray}
Note that, this (second order) IF of the MDPDE based Wald-type test statistic directly depends on $IF\big(y,T_{\alpha},{\boldsymbol{\theta}_{0}}\big)$,
the IF of the underlying MDPDE used. Based on our earlier exploration in Section \ref{SEC:MDPDE_IF}, 
the test IF in (\ref{EQ:IF2_test_Gen}) will then be bounded in the contamination point $y$ for any $\alpha>0$
which implies the robustness of the test based on the MDPDE based statistics in (\ref{EQ:TestStat_gen}).

Again from the general theory of \cite{Ghosh/etc:2016}, one can see the robustness of the level and power 
of the MDPDE based Wald-type tests for any $\alpha>0$ through their bounded level influence function (LIF)
and the power influence function (PIF). In particular, the LIF of any order is identically zero and
the PIF for testing at the significance level $\tau_0$ has the form  
\begin{eqnarray}
PIF\big(y, W_{\alpha, n}, {\boldsymbol{\theta}_{0}}\big) 
=C_{r}^{*}\left(\boldsymbol{d}^{T}\boldsymbol{Q}_\alpha\big(\boldsymbol{\theta}_{0}\big)\boldsymbol{d}\right)
\boldsymbol{d}^{T}\boldsymbol{Q}_\alpha\big(\boldsymbol{\theta}_{0}\big)IF(y, T_{\alpha}, {\boldsymbol{\theta}_{0}}),
\label{EQ:PIF_gen}
\end{eqnarray} 
where  
$C_{r}^{*}\big(s\big)=e^{-\frac{s}{2}}\sum_{v=0}^{\infty}{s^{v-1}}{2^{-v}}(2v-s)P\big(\chi^{2}_{r+2v}>\chi^{2}_{r,\tau_0}\big)/v!.
$
Again the PIF is a linear function of the IF of the MDPDE 
and hence bounded for all $\alpha>0$ indicating power robustness of the Wald-type test based on (\ref{EQ:TestStat_gen}).

\subsection{Robust Test for Symmetry}
\label{SEC:Test_Symmetry}

We now discuss, in detail, a particular testing problem in the context of SN distribution, namely the test of symmetry through the null hypothesis
$H_0: \gamma=0$. Let us consider a slightly general problem of testing
\begin{eqnarray}
H_{0} : \gamma =\gamma_{0} \hspace{2mm} \mbox{against} \hspace{2mm} 
H_{1}: \gamma \neq \gamma_{0},
\label{EQ:Hyp_gamma}
\end{eqnarray}
for a pre-fixed real $\gamma_{0}$. Note that, the choice $\gamma_{0}=0$ yields the test for symmetry against the SN alternatives.
Note that, here $\mu$ and $\sigma$ are unknown nuisance parameters.
In the notation of Section \ref{SEC:Test_Gen}, we have 
$\Theta_{0}=\big\{\boldsymbol{\theta}=(\mu,\sigma,\gamma)^T : \mu \in \mathbb{R}, \sigma \in \mathbb{R}^{+}, \gamma = 0 \big \}$,
$r=1$, $\boldsymbol{m}(\boldsymbol{\theta})=\gamma-\gamma_{0}$ and $M(\boldsymbol{\theta})= (0,0,1)^{T}$.

Denoting the MDPDE as $\widehat{\boldsymbol{\theta}}_{\alpha,n}=\big(\widehat{\mu}_{\alpha,n},\widehat{\sigma}_{\alpha,n},\widehat{\gamma}_{\alpha,n}\big)$,
our MDPDE based Wald-type test statistics (\ref{EQ:TestStat_gen}) has a simplified form for testing (\ref{EQ:Hyp_gamma})
which is given by 
\begin{eqnarray}
W_{\alpha,n}= \frac{n\left(\widehat{\gamma}_{\alpha,n} - \gamma_{0}\right)^{2}}{\Sigma_\alpha^{(33)}(\widehat{\boldsymbol{\theta}}_{\alpha,n})},
\label{EQ:TestStat_gamma}
\end{eqnarray}
where $\Sigma_\alpha^{(33)}(\boldsymbol{\theta})$ is the $(3,3)$-th element of $\boldsymbol{\Sigma}_\alpha(\boldsymbol{\theta})$. 
Then, under the null hypothesis in (\ref{EQ:Hyp_gamma}), $W_{\alpha,n}$ asymptotically follows $\chi_{1}^{2}$ distribution
and the test can be performed by comparing $W_{\alpha, n}$ with the corresponding critical values.
Further, the approximate expression of power function at the contiguous hypothesis of the form 
$H_{1,n}: \gamma= \gamma_{0}+ n^{-1/2}d$, with $d\in \mathbb{R}$,
is given by  
$$
\Pi_{\alpha}\big(\boldsymbol{\theta}_{n}\big)=1-G_{\chi_{1,\delta}^{2}}\big(\chi_{1,\tau_0}^{2}\big),
~~~~~\mbox{with}~~\delta=\frac{d^{2}}{\Sigma_\alpha^{(33)}(\boldsymbol{\theta}_0)}, ~\boldsymbol{\theta}_0\in\Theta_{0}.
$$
We have numerically calculated this asymptotic contiguous power for testing symmetry ($\gamma_{0}=0$) at 5\% level of significance
by the MDPDE based Wald-type test with different values of $\alpha$, which  are presented in Table \ref{TAB:Power}
for $\boldsymbol{\theta}_{0}=(0,1,0)^T$. It is clear that, just like the ARE of the MDPDE, 
the contiguous power of the MDPDE based test also decreases as $\alpha$ increases but this loss is not quite significant at small $\alpha>0$.
For larger values of $d$, i.e., alternatives further away from the null, the power eventually becomes one for all $\alpha\geq 0$
in accordance with the consistency of these tests.

\begin{table}[!h]
	\caption{Asymptotic contiguous power of the MDPDE based Wald-type test for testing symmetry ($\gamma_0=0$) at 5\% level of significance 
	with $\boldsymbol{\theta}_{0}=(0,1,0)^T$ and different values of $\alpha$ and $d$}
	\centering	
\begin{tabular}{c |c c c c c c c c}
	\hline
 & \multicolumn{8}{|c}{$\alpha$}\\
d & 0 & 0.05 & 0.1 & 0.2 & 0.3 & 0.5 & 0.7 & 1\\
	\hline
3.00 & 0.6685 & 0.6678 & 0.6662 & 0.6471 & 0.6327 & 0.6011 & 0.5507 & 0.4905\\
3.50 & 0.7982 & 0.7975 & 0.7960 & 0.7785 & 0.7649 & 0.7342 & 0.6827 & 0.6175\\
4.00 & 0.8915 & 0.8910 & 0.8899 & 0.8763 & 0.8655 & 0.8401 & 0.7948 & 0.7329\\
4.50 & 0.9489 & 0.9485 & 0.9478 & 0.9390 & 0.9317 & 0.9137 & 0.8792 & 0.8275\\
5.00 & 0.9790 & 0.9788 & 0.9784 & 0.9736 & 0.9694 & 0.9585 & 0.9356 & 0.8974\\
5.50 & 0.9925 & 0.9924 & 0.9922 & 0.9900 & 0.9879 & 0.9823 & 0.9690 & 0.9440\\
6.00 & 0.9977 & 0.9977 & 0.9976 & 0.9967 & 0.9958 & 0.9933 & 0.9866 & 0.9721\\
7.00 & 0.9999 & 0.9999 & 0.9998 & 0.9998 & 0.9997 & 0.9993 & 0.9982 & 0.9947\\
8.00 & 1.0000 & 1.0000 & 1.0000 & 1.0000 & 1.0000 & 1.0000 & 0.9998 & 0.9993\\
9.00 & 1.0000 & 1.0000 & 1.0000 & 1.0000 & 1.0000 & 1.0000 & 1.0000 & 0.9999\\
	\hline
\end{tabular}
	\label{TAB:Power}
\end{table}

\begin{figure}[!h]
	\centering
	\includegraphics[width=0.4\textwidth]{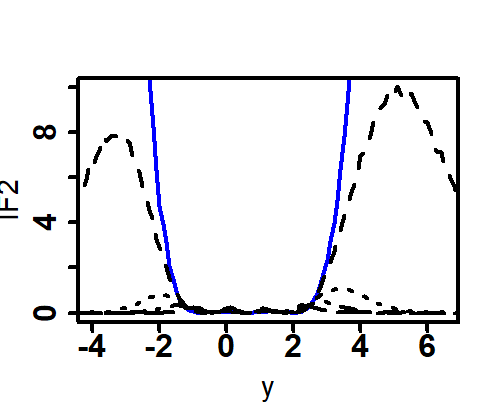}
	\includegraphics[width=0.4\textwidth]{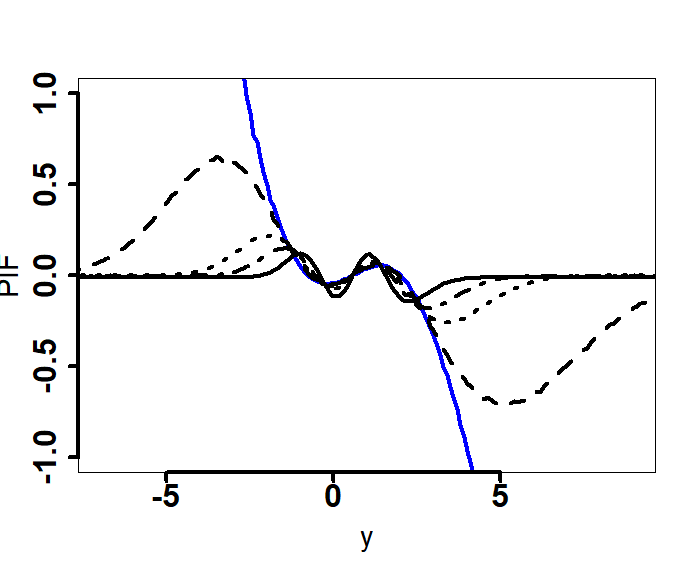}
\caption{Second Order IF and the PIF for the MDPDE based of Wald-type test for testing $\gamma = 1$
	with $\boldsymbol{\theta}_{0}=(0,1,1)^T$ and $d=4$, for different $\alpha$
	[blue solid line: $\alpha=0$ (MLE); dashed line: $\alpha=0.1$; dotted line: $\alpha=0.3$; dash-dotted line: $\alpha=0.5$; 
	black solid line: $\alpha=1$] }
\label{FIG:IF_test}
\end{figure}

Next the robustness of the Wald-type test based on the statistic (\ref{EQ:TestStat_gamma}) for testing (\ref{EQ:Hyp_gamma})
can be studied through the second order influence function of the test statistics and the PIF.
From the general formulas presented in Section \ref{SEC:Test_Gen}, we can easily calculate these measures 
in the present case of testing (\ref{EQ:Hyp_gamma}) as given by 
\vspace{-1cm}
\begin{eqnarray}
IF_{2}\big(y,W_{\alpha,n},{\boldsymbol{\theta}_{0}}\big)
&=&2 IF\big(y,T_{\alpha}^{(\gamma)},{\boldsymbol{\theta}_{0}}\big)^{2}/\Sigma_\alpha^{(33)}(\boldsymbol{\theta}_0).
\label{EQ:IF2_test_Gamma}
\\
PIF\big(y, W_{\alpha, n}, {\boldsymbol{\theta}_{0}}\big) 
&=&C_{r}^{*}\left({d}^{2}/\Sigma_\alpha^{(33)}(\boldsymbol{\theta}_0)\right)
IF(y, T_{\alpha}^{(\gamma)}, {\boldsymbol{\theta}_{0}})d/\Sigma_\alpha^{(33)}(\boldsymbol{\theta}_0),
\label{EQ:PIF_gamma}
\end{eqnarray} 
where $T_{\alpha}^{(\gamma)}$ is the MDPDE functional corresponding to $\gamma$ and 
hence $IF(y, T_{\alpha}^{(\gamma)}, {\boldsymbol{\theta}_{0}})$ is given by 
the third component of the full 3-dimensional IF vector  given in (\ref{EQ:MDPDE_IF}).
For illustration, we have presented the plots of these $IF_2$ and $PIF$ for different values of $\alpha$
at $\gamma_0=1$, $\boldsymbol{\theta}_{0}=(0,1,1)^T$ and $d=4$ in Figure \ref{FIG:IF_test};
note that the plot of the corresponding $IF(y, T_{\alpha}^{(\gamma)}, {\boldsymbol{\theta}_{0}})$ 
is as given in Figure \ref{FIG:IF_gamma}.
It is clearly evident from these figures that 
the MDPDE based Wald-type test statistic (\ref{EQ:TestStat_gamma}) has bounded second order IF 
as well as bounded PIF for all $\alpha>0$
indicating the claimed robustness and the extent of robustness increases as $\alpha$ increases.

\section{Simulation study}
\label{SEC:Simulation}

\subsection{Performance of the MDPDE}
\label{SEC:Simulation_MDPDE}

We now examine the finite sample performances of the MDPDE for the SN distribution through a Monte-Carlo simulation study. 
We simulate random samples from the SN distribution, using the R package '\textit{Sn}', 
for different sizes $n=50$, 100, and the true parameter values $(\mu,\sigma, \gamma)=(0,1,5)$.
Based on each simulated sample, we compute the MDPDEs at different $\alpha$, including the MLE at $\alpha=0$, 
using the genetic algorithm described in Section \ref{SEC:MDPDE_Computation}.
Replicating this process 500 times, we compute the empirical bias and MSE of 
the MDPDEs of the three parameters $(\mu,\sigma, \gamma)$ for  the fitted SN distribution.
Further, to examine the robustness property, we repeat the above simulation exercise 
by  contaminating  $100\epsilon\%$ of each sample by observations from a distant contaminating distribution.
We have considered the contamination proportion $\epsilon$ to be  0.05 and 0.1,
leading to 5\% and 10\% contaminations respectively, and four different contaminating distributions as 
SN(10,1,5), SN($-10$,1,5), SN(0,5,5) and SN(0,1,1).
For each situation, the bias and MSE under contaminated data are again computed using 500 replications 
for MDPDEs at different $\alpha$ and are compared with their pure data values. 
We report the exact values of the biases and MSEs of each of the  three parameters $(\mu,\sigma, \gamma)$ 
under each scenario in Table \ref{TAB:Sim_n100}.

\begin{table}[!h]
	\caption{Empirical Biases and MSEs of the MDPDEs for various $\alpha$ at sample size $n=50$}
	\centering	
	\resizebox{\textwidth}{!}{
		\begin{tabular}{ c c| c c c c c c| c c c c c c}
			\hline
			& & & & $\alpha$ & & &  & & $\alpha$ & & &\\
			$\epsilon$ &  & 0(MLE) & 0.1 & 0.3 & 0.5 & 0.7 & 1 & 0(MLE) & 0.1 & 0.3 & 0.5 & 0.7 & 1\\
			\hline
			&&		\multicolumn{6}{c|}{Bias}&		\multicolumn{6}{|c}{MSE}\\
			\hline
			0 &  $\mu$ & 0.009 & 0.015 & 0.017 & 0.023 & 0.050 & 0.100
			& 0.006 & 0.008 & 0.010 & 0.012 & 0.019 & 0.028\\
			& $\sigma$ & -0.012 & -0.086 & -0.100 & -0.112 & -0.186 & -0.253
			& 0.008 & 0.017 & 0.023 & 0.040 & 0.096 & 0.118\\
			& $\gamma$ & 0.908 & 1.028 & 1.094 & 1.207 & 1.302 & 1.376
			& 1.718 & 2.178 & 2.768 & 3.024 & 3.572 & 3.910\\
			\hline
			\multicolumn{14}{l}{\textit{Outliers from SN(10,1,5)}}\\
			0.05 & $\mu$ & 0.124 & 0.096 & 0.090 & 0.048 & 0.050 & 0.116
			& 0.038 & 0.024 & 0.020 & 0.015 & 0.026 & 0.041\\
			& $\sigma$ & -0.316 & -0.296 & -0.177 & -0.116 & -0.170 & -0.278
			& 0.118 & 0.102 & 0.072 & 0.042 & 0.101 & 0.133\\
			& $\gamma$ & 1.229 & 1.169 & 1.088 & 0.992 & 1.276 & 1.398
			& 3.745 & 3.272 & 3.129 & 3.049 & 3.802 & 4.172\\
			0.1 & $\mu$ & 0.130 & 0.117 & 0.109 & 0.059 & 0.066 & 0.174
			& 0.041 & 0.030 & 0.028 & 0.017 & 0.039 & 0.079\\
			& $\sigma$ & -0.373 & -0.327 & -0.226 & -0.126 & -0.177 & -0.302
			& 0.158 & 0.141 & 0.092 & 0.053 & 0.151 & 0.170\\
			& $\gamma$ & 1.411 & 1.338 & 1.185 & 1.070 & 1.342 & 1.446
			& 3.958 & 3.573 & 3.385 & 3.128 & 3.990 & 4.360\\
			\hline	
			\multicolumn{14}{l}{\textit{Outliers from SN(-10,1,5)}}\\
			0.05 & $\mu$ & -0.145 & -0.098 & -0.084 & 0.047 & 0.080 & 0.117
			& 0.046 & 0.027 & 0.021 & 0.016 & 0.024 & 0.044\\
			& $\sigma$ & -0.272 & -0.236 & -0.121 & -0.108 & -0.178 & -0.310
			& 0.171 & 0.113 & 0.079 & 0.043 & 0.104 & 0.139\\
			& $\gamma$ & 1.217 & 1.138 & 1.081 & 0.961 & 1.192 & 1.368
			& 3.698 & 3.240 & 3.119 & 3.029 & 3.598 & 4.062\\
			0.1 & $\mu$ & -0.154 & -0.114 & -0.095 & 0.068 & 0.092 & 0.126
			& 0.052 & 0.034 & 0.029 & 0.018 & 0.034 & 0.049\\
			& $\sigma$ &-0.327 & -0.277 & -0.122 & -0.110 & -0.182 & -0.314
			& 0.233 & 0.141 & 0.094 & 0.052 & 0.115 & 0.182\\
			& $\gamma$ & 1.285 & 1.149 & 1.093 & 1.008 & 1.230 & 1.425
			& 3.786 & 3.470 & 3.218 & 3.096 & 3.996 & 4.281\\
			\hline
			\multicolumn{14}{l}{\textit{Outliers from SN(0,5,5)}}\\
			0.05 & $\mu$ & 0.090 & 0.079 & 0.064 & 0.063 & 0.086 & 0.122
			& 0.036 & 0.024 & 0.019 & 0.014 & 0.023 & 0.029\\
			& $\sigma$ & 0.291 & 0.203 & 0.142 & 0.113 & -0.170 & -0.286
			& 0.334 & 0.269 & 0.168 & 0.095 & 0.138 & 0.218\\
			& $\gamma$ & 1.270 & 1.126 & 1.058 & 0.971 & 1.265 & 1.316
			& 3.841 & 3.259 & 3.134 & 2.977 & 3.635 & 4.014\\
			0.1 & $\mu$ & 0.105 & 0.087 & 0.073 & 0.071 & 0.094 & 0.144
			& 0.051 & 0.041 & 0.028 & 0.020 & 0.023 & 0.035\\
			& $\sigma$ & 0.336 & 0.298 & 0.151 & -0.124 & -0.214 & -0.317
			& 0.413 & 0.341 & 0.236 & 0.124 & 0.210 & 0.295\\
			& $\gamma$ & 1.324 & 1.168 & 1.094 & 1.003 & 1.337 & 1.409
			& 3.927 & 3.478 & 3.197 & 3.086 & 3.942 & 4.388\\
			\hline
			\multicolumn{14}{l}{\textit{Outliers from SN(0,1,1)}} \\
			0.05 & $\mu$ & -0.072 & -0.058 & -0.043 & 0.036 & 0.080 & 0.122
			& 0.094 & 0.040 & 0.021 & 0.014 & 0.022 & 0.031\\
			& $\sigma$ & -0.257 & -0.171 & -0.146 & -0.115 & 0.168 & 0.292
			& 0.225 & 0.197 & 0.133 & 0.073 & 0.108 & 0.164\\
			& $\gamma$ & 1.192 & 1.148 & 1.115 & 1.088 & 1.178 & 1.418
			& 3.736 & 3.433 & 3.290 & 3.127 & 3.263 & 4.273\\
			0.1 & $\mu$ & -0.091 & -0.076 & 0.074 & 0.063 & 0.107 & 0.132
			& 0.130 & 0.084 & 0.051 & 0.020 & 0.028 & 0.051\\
			& $\sigma$ & -0.299 & -0.254 & -0.192 & 0.150 & 0.187 & 0.325
			& 0.332 & 0.245 & 0.140 & 0.089 & 0.124 & 0.195\\
			& $\gamma$ & 1.325 & 1.256 & 1.132 & 1.102 & 1.232 & 1.486
			& 3.949 & 3.595 & 3.420 & 3.146 & 3.642 & 4.447\\
			\hline
		\end{tabular}
	}
	\label{TAB:Sim_n50}
\end{table}

\begin{table}[!h]
	\caption{Empirical Biases and MSEs of the MDPDEs for various $\alpha$ at sample size $n=100$}
	\centering	
	\resizebox{\textwidth}{!}{
		\begin{tabular}{ c c| c c c c c c| c c c c c c}
			\hline
			& & & & $\alpha$ & & &  & & $\alpha$ & & &\\
			$\epsilon$ &  & 0(MLE) & 0.1 & 0.3 & 0.5 & 0.7 & 1& 0(MLE) & 0.1 & 0.3 & 0.5 & 0.7 & 1\\
			\hline
			&&		\multicolumn{6}{c|}{Bias}&		\multicolumn{6}{|c}{MSE}\\
			\hline
			 0 &  $\mu$ & 0.006 & 0.014 & 0.016 & 0.023 & 0.032 & 0.084
			& 0.004 & 0.004 & 0.005 & 0.006 & 0.007 & 0.011\\
			 & $\sigma$ & -0.004 & -0.076 & -0.091 & -0.105 & -0.154 & -0.209
			& 0.006 & 0.007 & 0.009 & 0.020 & 0.038 & 0.063\\
			 & $\gamma$ & 0.495 & 0.563 & 0.781 & 1.016 & 1.142 & 1.201
			& 0.912 & 1.182 & 1.334 & 2.193 & 3.020 & 3.223\\
			\hline
			\multicolumn{14}{l}{\textit{Outliers from SN(10,1,5)}}\\
			 0.05 & $\mu$ & 0.102 & 0.087 & 0.060 & 0.024 & 0.042 & 0.095
			& 0.028 & 0.020 & 0.016 & 0.008 & 0.014 & 0.034\\
			 & $\sigma$ & -0.260 & -0.223 & -0.118 & -0.105 & -0.158 & -0.215
			& 0.098 & 0.064 & 0.049 & 0.027 & 0.100 & 0.130\\
			 & $\gamma$ & 1.100 & 1.038 & 1.010 & 0.945 & 1.150 & 1.244
			& 3.027 & 2.814 & 2.408 & 2.273 & 3.099 & 3.542\\
			 0.1 & $\mu$ & 0.103 & 0.089 & 0.064 & 0.028 & 0.050 & 0.105
			& 0.032 & 0.022 & 0.018 & 0.010 & 0.016 & 0.042\\
			 & $\sigma$ & -0.338 & -0.257 & -0.145 & -0.109 & -0.168 & -0.258
			& 0.141 & 0.092 & 0.066 & 0.035 & 0.116 & 0.150\\
			 & $\gamma$ & 1.108 & 1.046 & 1.010 & 0.999 & 1.194 & 1.341
			& 3.187 & 3.038 & 2.992 & 2.658 & 3.296 & 3.962\\
			\hline	
			\multicolumn{14}{l}{\textit{Outliers from SN(-10,1,5)}}\\
			 0.05 & $\mu$ & -0.126 & -0.085 & 0.057 & 0.024 & 0.039 & 0.101
			& 0.035 & 0.021 & 0.015 & 0.008 & 0.015 & 0.033\\
			 & $\sigma$ & -0.221 & -0.191 & -0.115 & -0.065 & -0.167 & -0.265
			& 0.138 & 0.086 & 0.045 & 0.021 & 0.100 & 0.126\\
			 & $\gamma$ & 1.132 & 1.044 & 1.011 & 0.955 & 1.163 & 1.229
			& 3.353 & 2.881 & 2.390 & 2.274 & 3.091 & 3.296\\
			 0.1 & $\mu$ & -0.142 & -0.096 & -0.063 & 0.031 & 0.057 & 0.111
			& 0.042 & 0.028 & 0.017 & 0.011 & 0.019 & 0.040\\
			 & $\sigma$ & -0.275 & -0.214 & -0.119 & -0.075 & -0.173 & -0.291
			& 0.196 & 0.098 & 0.058 & 0.032 & 0.104 & 0.164\\
			 & $\gamma$ & 1.187 & 1.096 & 1.072 & 0.986 & 1.189 & 1.260
			& 3.598 & 3.072 & 2.890 & 2.580 & 3.458 & 3.911\\
			\hline
			\multicolumn{14}{l}{\textit{Outliers from SN(0,5,5)}}\\
			 0.05 & $\mu$ & 0.070 & 0.064 & 0.053 & 0.047 & 0.085 & 0.108
			& 0.018 & 0.010 & 0.009 & 0.006 & 0.010 & 0.018\\
			& $\sigma$ & 0.233 & 0.168 & 0.118 & 0.105 & -0.147 & -0.231
			& 0.299 & 0.193 & 0.156 & 0.078 & 0.117 & 0.196\\
			& $\gamma$ & 1.168 & 1.051 & 1.007 & 0.861 & 1.154 & 1.206
			& 3.164 & 2.881 & 2.465 & 2.243 & 3.172 & 3.805\\
			0.1 & $\mu$ & 0.077 & 0.072 & 0.057 & 0.050 & 0.091 & 0.118
			& 0.039 & 0.028 & 0.016 & 0.009 & 0.019 & 0.024\\
			& $\sigma$ & 0.264 & 0.240 & 0.136 & 0.117 & 0.184 & -0.230
			& 0.378 & 0.253 & 0.185 & 0.098 & 0.181 & 0.262\\
			& $\gamma$ & 1.235 & 1.100 & 1.069 & 0.974 & 1.270 & 1.349
			& 3.499 & 3.132 & 2.987 & 2.567 & 3.580 & 4.068\\
			\hline
			\multicolumn{14}{l}{\textit{Outliers from SN(0,1,1)}} \\
			 0.05 & $\mu$ & -0.057 & -0.053 & 0.039 & 0.028 & 0.068 & 0.092
			& 0.015 & 0.011 & 0.008 & 0.007 & 0.010 & 0.013\\
			 & $\sigma$ & -0.194 & -0.133 & -0.124 & 0.113 & 0.126 & 0.216
			& 0.166 & 0.103 & 0.090 & 0.044 & 0.062 & 0.133\\
			& $\gamma$ & 1.106 & 1.049 & 1.012 & 1.002 & 1.095 & 1.266
			& 3.040 & 2.832 & 2.398 & 2.278 & 3.196 & 3.622\\
			0.1 & $\mu$ & -0.065 & -0.064 & 0.053 & 0.051 & 0.074 & 0.106
			& 0.033 & 0.029 & 0.010 & 0.009 & 0.016 & 0.019\\
			& $\sigma$ & -0.256 & -0.149 & 0.137 & 0.118 & 0.142 & 0.274
			& 0.203 & 0.178 & 0.122 & 0.075 & 0.096 & 0.158\\
			& $\gamma$ & 1.230 & 1.188 & 1.087 & 1.020 & 1.130 & 1.292
			& 3.391 & 3.098 & 2.781 & 2.398 & 3.305 & 4.010\\
			\hline
		\end{tabular}
	}
	\label{TAB:Sim_n100}
\end{table}

One can clearly note that the bias and MSE under pure data increases with $\alpha$ 
but the increase is reasonably smaller at smaller positive values of $\alpha$.
On the other hand, under contaminated data the bias and MSE increases significantly for the MLE (at $\alpha=0$), 
whereas those for MDPDEs with larger $\alpha$ remains more closer to their pure data values;
the stability increases with increasing values of $\alpha>0$.
Based on the efficiency and robustness trade-off, it has been observed in all the situations considered,
the MDPDE with $\alpha$ around 0.5 produce smallest values of bias and MSEs under contamination
which are significantly lower compared to those obtained by the MLE under contamination.

\subsection{Performance of the MDPDE based Wald-type test}
\label{SEC:Simulation_test}

To visualize the performance of proposed MDPDE based Wald-type tests,
we have again performed several simulation studies. 
We consider the problem of testing symmetry through the hypothesis $H_{0}$: $\gamma=0$ against $H_{1}$: $\gamma \neq 0$,
for which the Wald-type test statistic $W_{\alpha,n}$ is as given in (\ref{EQ:TestStat_gamma}) with $\gamma_0=0$. 
We first simulate random samples of sizes $n=50,100$ from the SN(0,1,0) distribution and 
perform the MDPDE based Wald-type test for different $\alpha$, including the classical Wald test at $\alpha=0$.
Based on 500 replications, we then compute the empirical levels of the tests measured as 
the proportion of test statistics exceeding the chi-square critical value among the 500 replications.
Subsequently, to compute the empirical power of the tests, we repeat the above exercise 
but now generating random samples from an alternative SN(0,1,1) distribution. 
Finally, to illustrate the claimed robustness, we recalculate the level and power of the Wald-type tests
after contamination $100\epsilon\%$ of each sample in the previous simulation exercises with $\epsilon=0.05, 0.1$.
The contaminated observations are generated from SN(0,1,3) and SN(0,1,$-$3) distributions, (?????)
respectively, for the level and power calculations. 
In Table \ref{TAB:Simulation_test}, we report all the resulting empirical levels and powers 
obtained from different simulation scenarios.

\begin{table}[!h]
	\caption{Empirical Levels and Powers of the MDPDE based Wald-type tests
	for different $\alpha$}
	\centering	
	\resizebox{0.7\textwidth}{!}{
	\begin{tabular}{c c c|c c c c c c}
		\hline
&		Sample  & & & & $\alpha$ & & &\\
&		Size ($n$) & $\epsilon$  & 0(MLE) & 0.1 & 0.3 & 0.5 & 0.7 & 1.0\\
		\hline
Level & 50 & 0 &0.12 & 0.144 & 0.162 & 0.188 & 0.22 & 0.25\\
	&	& 0.05 & 0.796 & 0.44 & 0.25 & 0.194 & 0.168 & 0.126\\
	&	& 0.10 & 0.848 & 0.486 & 0.282 & 0.216 & 0.184 & 0.14\\\\
	&100 & 0  & 0.058 & 0.102 & 0.124 & 0.168 & 0.18 & 0.202\\
	&	& 0.05 & 0.862 & 0.52 & 0.264 & 0.19 & 0.134 & 0.116\\
	&	& 0.10 & 0.876 & 0.55 & 0.296 & 0.202 & 0.148 & 0.124\\
		\hline
Power	& 50 & 0 & 0.95 & 0.962 & 0.978 & 0.99 & 0.998 & 1.0\\
		& & 0.05 & 0.24 & 0.57 & 0.836 & 0.986 & 1 & 1\\
		& & 0.10 & 0.254 & 0.582 & 0.854 & 0.992 & 1 & 1\\\\
	& 100 & 0 & 0.974 & 0.98 & 0.99 & 1 & 1 & 1\\
		& & 0.05 & 0.32 & 0.632 & 0.884 & 0.99 & 1 & 1\\
		& & 0.10 & 0.342 & 0.676 & 0.902 & 1 & 1 & 1\\
		\hline
	\end{tabular}}
\label{TAB:Simulation_test}
\end{table}

It can be observed from the table that, under pure data, the levels are inflated for Wald-type tests with larger $\alpha$.
However, through more extensive simulations (not presented here for brevity) that 
the levels stabilizes to the desired 5\% significance level for larger sample sizes;
although this happens for the classical MLE based Wald test (at $\alpha=0$) at $n=100$ itself, 
it needs much larger sample sizes to achieve desired level for larger values of $\alpha$.
As a results, the pure data power always appears higher for the Wald-type tests with larger $\alpha$ 
and they indeed becomes one  for all $\alpha$ at moderately large sample sizes. 
However, the main advantage of the MDPDE based Wald-type tests appear at 
the stability of their levels and sizes under contamination in sample data.
For Wald test at $\alpha=0$, the level inflates significantly due to contamination
but becomes more stable with increasing $\alpha$.
Similarly, the power of the classical Wald test decreases drastically under contamination
but regain its high values  for the MDPDE based Wald-type tests with larger $\alpha\geq 0.3$.
Therefore the MDPDE based Wald-type tests with moderately large $\alpha>0$ always produce more power 
with a slightly inflated levels which remain stable even under different contamination  levels.

\section{Real Data Applications}

\subsection{AIS Dataset}

Let us consider again the motivating dataset and use the MDPDE to obtain the estimates of the fitted SN distributions.
We consider again the important health indicator variables as in Figure \ref{FIG:Data_hist}
and compute the MDPDEs of the parameters of the fitted SN distribution for each variable
using the algorithm described in Section \ref{SEC:MDPDE_Computation}.
We have also estimated the standard errors of the resulting MDPDEs using the formula described in Section \ref{SEC:MDPDE_Asymp}.
The parameter estimates, along with their standard errors, for all eight variables are reported in Table \ref{TAB:Data_MDPDE}.
The outlier deleted MLE, obtained after removing the outliers identified through the respective box-plots,
are also presented in Table \ref{TAB:Data_MDPDE} for reference.

\begin{table}
	\caption{The SN parameter estimates (standard errors) for the measurements in AIS data, 
		obtained through MDPDEs at different $\alpha$, 
		the MLE (at $\alpha=0$) and the outlier deleted MLE.
	The number of outliers found for each measurements are reported after their name in the first column.}
	\centering	
	\resizebox{0.8\textwidth}{!}{
\begin{tabular}{c c|c c c c c c| c}
	\hline
	Variable &  & & & & $\alpha$ & & & Outlier deleted \\
	(Outlier)&  & 0(MLE) & 0.1 & 0.3 & 0.5 & 0.7 & 1.0 & MLE\\
	\hline
	HC & $\mu$ & 40.664 & 43.387 & 45.384 & 46.382 & 46.382 & 46.383 & 46.440\\
	(1)& & (0.187) & (0.339) & (0.546) & (0.670) & (1.084) & (1.463) & (0.336)\\
	& $\sigma$ & 4.387 & 4.883 & 4.881 & 4.876 & 4.876 & 4.876 & 4.880\\
	& & (0.185) & (0.330) & (0.438) & (0.641) & (1.300) & (1.802) & (0.333)\\
	& $\gamma$ & 0.966 & -1.752 & -1.762 & -1.766 & -1.766 & -1.766 & -1.794\\
	& & (0.104) & (0.472) & (0.661) & (0.942) & (1.354) & (1.979) & (0.449)\\
	\hline
	RCC & $\mu$ & 4.296 & 4.543 & 4.539 & 4.525 & 4.528 & 4.529 & 4.543 \\
	(1)& & (0.020) & (0.021) & (0.022) & (0.024) & (0.027) & (0.028) & (0.018)\\
	& $\sigma$ & 0.622 & 0.525 & 0.530 & 0.504 & 0.521 & 0.533 & 0.466\\
	& & (0.024) & (0.026) & (0.027) & (0.028) & (0.031) & (0.035) & (0.022)\\
	& $\gamma$ & 1.607 & 0.506 & 0.502 & 0.500 & 0.499 & 0.498 & 0.499\\
	& & (0.086) & (0.090) & (0.095) & (0.100) & (0.107) & (0.153) & (0.086)\\
	\hline
	WCC & $\mu$ & 5.106 & 5.481 & 5.472 & 5.471 & 5.471 & 5.471 & 5.475\\
	(4)& & (0.098) & (0.099) & (0.105) & (0.106) & (0.119) & (0.156) & (0.094)\\
	& $\sigma$ & 2.690 & 2.311 & 2.197 & 2.193 & 2.191 & 2.191 & 2.184\\
	& & (0.105) & (0.109) & (0.125) & (0.126) & (0.147) & (0.196) & (0.099)\\
	& $\gamma$ & 2.727 & 1.716 & 1.712 & 1.712 & 1.712 & 1.712 & 1.703\\
	& & (0.146) & (0.164) & (0.164) & (0.166) & (0.216) & (0.250) & (0.143)\\
	\hline
	PFC & $\mu$ & 20.244 & 23.197 & 23.196 & 23.196 & 23.196 & 23.196 & 23.226\\
	(12)& & (1.519) & (1.595) & (1.742) & (1.834) & (1.977) & (2.200) & (1.546)\\
	& $\sigma$ & 73.840 & 67.835 & 61.832 & 57.832 & 57.832 & 57.832 & 57.671\\
	& & (2.905) & (3.178) & (3.663) & (3.689) & (4.386) & (4.868) & (2.934)\\
	& $\gamma$ & 9.143 & 7.096 & 6.097 & 6.097 & 6.097 &  6.097 & 6.066\\
	& & (0.361) & (0.606) & (0.628) & (0.854) & (1.097) & (1.783) & (0.402)\\
	\hline
	BMI & $\mu$ & 19.970 & 21.344 & 22.294 & 22.291 & 22.291 & 22.291 & 22.315\\
	(7)& & (0.066) & (0.067) & (0.072) & (0.077) & (0.079) & (0.169) & (0.060)\\
	& $\sigma$ & 4.133 & 2.646 & 2.386 & 2.369 & 2.368 & 2.369 & 2.349\\
	& & (0.125) & (0.132) & (0.136) & (0.146) & (0.159) & (0.190) & (0.117)\\
	& $\gamma$ & 2.313 & 1.227 & 0.595 & 0.194 & 0.195 & 0.194 & 0.174\\
	& & (0.084) & (0.087) & (0.092) & (0.096) & (0.105) & (0.216) & (0.086)\\
	\hline
	LBM & $\mu$ & 50.383 & 50.953 & 50.953 & 50.953 & 50.953 & 50.953 & 50.958\\
	(1)& & (0.765) & (0.796) & (0.801) & (0.854) & (0.959) & (1.351) & (0.768)\\
	& $\sigma$ & 19.493 & 18.726 & 18.727 & 18.726 & 18.726 & 18.726 & 18.718\\
	& & (0.856) & (0.886) & (0.955) & (1.124) & (1.349) & (1.978) & (0.840)\\
	& $\gamma$ & 2.424 & 2.197 & 2.197 & 2.197 & 2.197 & 2.197 & 2.195\\
	& & (0.106) & (0.199) & (0.199) & (0.212) & (0.229) & (0.531) & (0.106)\\
	\hline
	Ht & $\mu$ & 187.072 & 184.794 & 184.794 & 184.794 & 184.794 & 184.794 & 184.771\\
	(3)& & (0.582) & (0.612) & (0.623) & (0.647) & (0.653) & (1.095) & (0.571)\\
	& $\sigma$ & 11.952 & 10.115 & 10.113 & 10.113 & 10.112 & 10.112 & 10.094\\
	& & (0.321) & (0.339) & (0.384) & (0.435) & (0.517) & (0.517) & (0.315)\\
	& $\gamma$ & -1.074 & -0.673 & -0.676 & -0.677 & -0.678 & -0.678 & -0.674\\
	& & (0.154) & (0.166) & (0.178) & (0.185) & (0.195) & (0.284) & (0.153)\\
	\hline
	Wt & $\mu$ & 64.066 & 69.063 & 72.063 & 72.063 & 72.063 & 72.062 & 72.143\\
	(4) & & (0.378) & (0.395) & (0.416) & (0.437) & (0.495) & (0.758) & (0.373)\\
	& $\sigma$ & 17.682 & 14.076 & 13.072 & 13.071 & 13.071 & 13.071 & 13.024\\
	& & (0.646) & (0.686) & (0.741) & (0.748) & (0.805) & (0.874) & (0.645)\\
	& $\gamma$ & 1.232 & 0.848 & 0.450 & 0.250 & 0.250 & 0.250 & 0.240\\
	& & (0.085) & (0.087) & (0.092) & (0.097) & (0.104) & (0.120) & (0.085)\\
	\hline
\end{tabular}}
\label{TAB:Data_MDPDE}
\end{table}

It can be easily observed from Table \ref{TAB:Data_MDPDE} that the MLE changes drastically for all the variables 
due to the presence of outliers, but the proposed MDPDEs with larger $\alpha>0$ computed over the full data 
remain extremely close to the outlier deleted MLE. Thus, the use of the MDPDEs with larger $\alpha>0$ 
leads to robust insights even in the presence of outliers in the data; 
most of the time the MDPDEs with large values of $\alpha$ are very close to the cleaned data MLE.  
However, we need larger values of $\alpha>0$ if the strength of the outliers increases (more in number or greater distance from the data center) 
and vice versa. In the present example,  the variables HC, RCC and LBM all have one outlying data-point
but the MDPDEs of RCC and LBM becomes quite close to the outlier deleted MLE at $\alpha\approx 0.1$
and that requires larger $\alpha\approx 0.5$ for HC due to the greater distance of the outlier in this case;
among other variables PFC, having 12 outliers, requires $\alpha \approx 0.5$ to generate robust estimates,
whereas the corresponding values of $\alpha$ are 0.3 also for the measurements BMI, WCC and Wt.  
In summary, all MDPDES with $\alpha\geq 0.5$ generates estimates similar to the outlier deleted MLE in all cases
although sometimes a substantially lower $\alpha>0$ may also produce stable results (like $\alpha=0.1$ for Ht).

\begin{figure}
	\centering
	\subfloat[Hemoglobin Conc.~(HC)]{
			\includegraphics[width=0.38\textwidth]{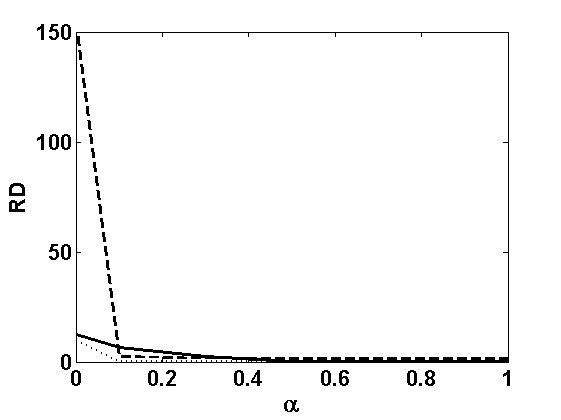}
	}
	\subfloat[Red Cell Count (RCC)]{
			\includegraphics[width=0.38\textwidth]{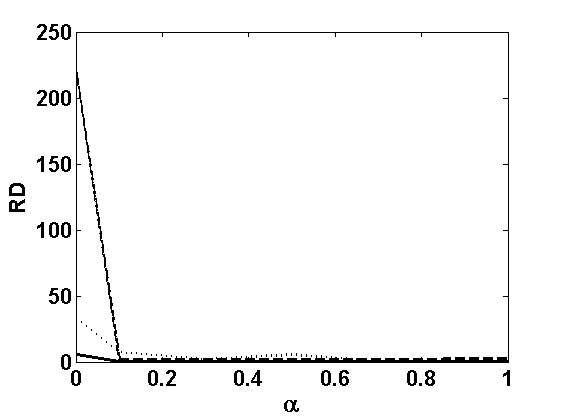}
	}
\\	
	\subfloat[White Cell Count (WCC)]{
			\includegraphics[width=0.38\textwidth]{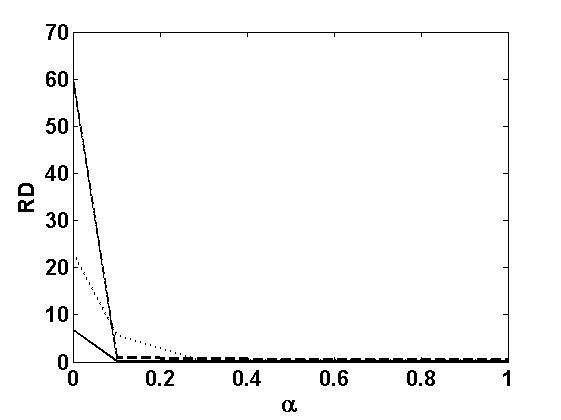}
	}
	\subfloat[Plasma Fer.~Conc.~(PFC)]{
			\includegraphics[width=0.38\textwidth]{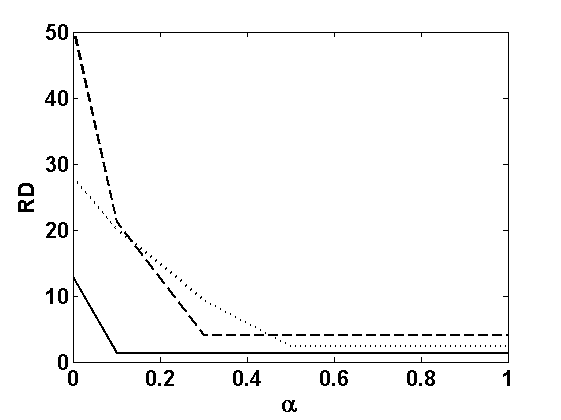}
	}
	\\
	\subfloat[Body Mass Index (BMI)]{
			\includegraphics[width=0.38\textwidth]{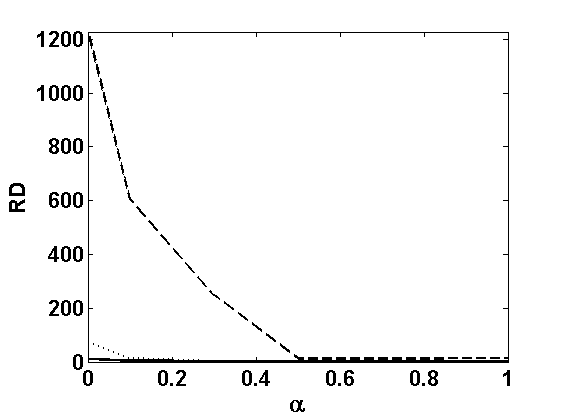}
	}
	\subfloat[Lean Body Mass (LBM)]{
			\includegraphics[width=0.38\textwidth]{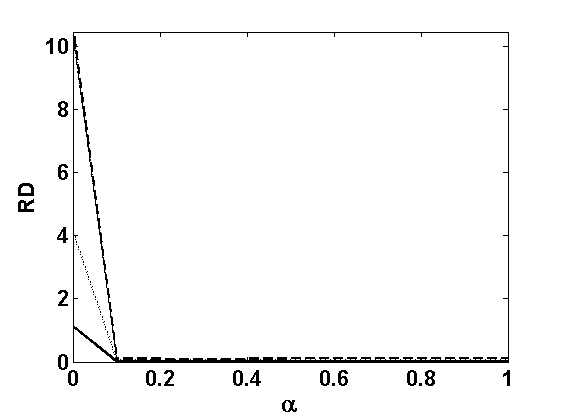}
	}
\\	
	\subfloat[Height (Ht)]{
			\includegraphics[width=0.38\textwidth]{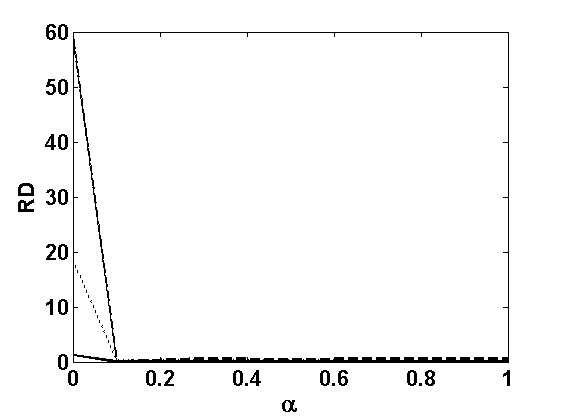}
	}
	\subfloat[Weight (Wt)]{
			\includegraphics[width=0.38\textwidth]{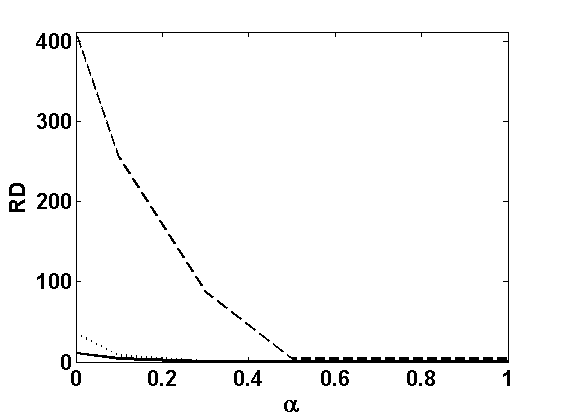}
	}	
	\caption{Relative differences (RDs) of the MDPDEs and MLE (at $\alpha=0$) due to the presence of outliers,
		plotted over $\alpha$, for different Health indicator variables from the AIS Data 
	[Solid line: $\mu$; dotted line: $\sigma$; dashed line: $\gamma$]}
	\label{FIG:Data_RD}
\end{figure}

To illustrate the robustness aspect of the MDPDEs more clearly, 
we have also recomputed the MDPDEs for outlier deleted data for all $\alpha\geq 0$
and compared them with the corresponding full data values; the greater robustness can be measured by the lower values  
of their relative differences defined as 
$$
RD(\nu)=\frac{|\widehat{\nu}_{\mbox{full}}- \widehat{\nu}_{\mbox{clean}}|}{\widehat{\nu}_{\mbox{full}}}\times 100\%,
~~~~~~~\nu \in\{\mu, \sigma, \gamma\},
$$
where $\widehat{\nu}_{\mbox{full}}$ and $\widehat{\nu}_{\mbox{clean}}$ denote, respectively, 
the estimates of $\nu\in\{\mu, \sigma, \gamma\}$ obtained from full data with outliers and the outlier deleted data.
For all the eight measurements, the relative differences (RDs) of the MDPDEs over different $\alpha$ are plotted in Figure \ref{FIG:Data_RD}.
Clearly the RDs are significantly high for MLE (at $\alpha=0$); they are as high as 1200\% and 400\% for 
the skewness parameter $\gamma$ for BMI and Wt, respectively.
But these RDs decrease for MDPDEs as $\alpha>0$ increases and become very close to zero for $\alpha\geq 0.5$ in all the cases;
they already become close to zero at $\alpha\approx 0.2$ for HC, RCC, WCC, LBM and Ht.
Among three parameters, the effect of outliers is seen to be most significant for $\gamma$ followed by $\sigma$
and the effect is often minimum for the parameter $\mu$. 
All these illustrations clearly show the claimed robustness of the proposed MDPDE 
with larger $\alpha>0$ for analyses of the present AIS dataset.

Next, let us study the performance of the proposed MDPDE based Wald-type tests for generating inference for the present AIS data.
We have examined several types of simple and composite parametric hypotheses for different health measurements in AIS data
with or without outliers. Since the results are similar in all cases, for brevity, 
we report the six most interesting cases as follows.
All six null hypotheses are composite as we assume the other remaining parameters to be unknown 
under the null as well as the corresponding omnibus alternative hypothesis.

\begin{figure}[!h]
	\centering
	\subfloat[Variable: HC; ~~$H_0 : \gamma=-1.8$]{
		\includegraphics[width=0.3\textwidth]{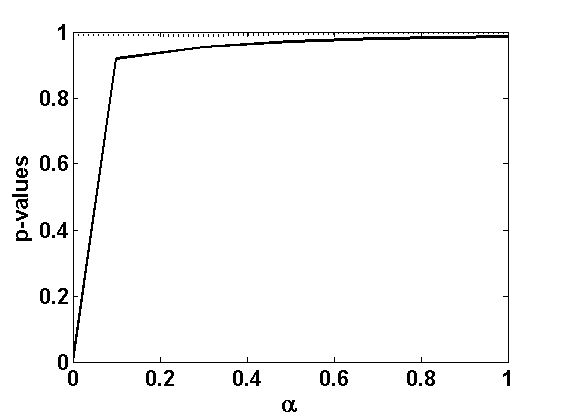}
	}
	\subfloat[Variable: WCC; ~~$H_0 : \gamma=1.7$]{
		\includegraphics[width=0.3\textwidth]{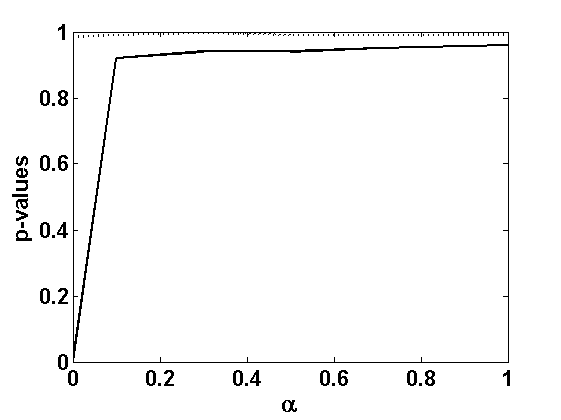}
	}
	\subfloat[Variable: LBM; ~~$H_0 : \gamma=2$]{
		\includegraphics[width=0.3\textwidth]{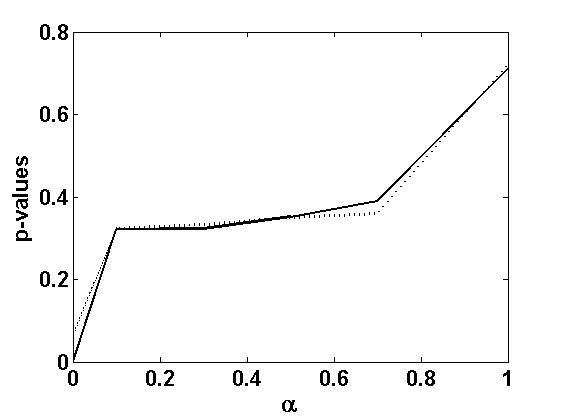}
	}
	\\
	\subfloat[Variable: PFC; ~~$H_0 : \sigma=57$]{
		\includegraphics[width=0.3\textwidth]{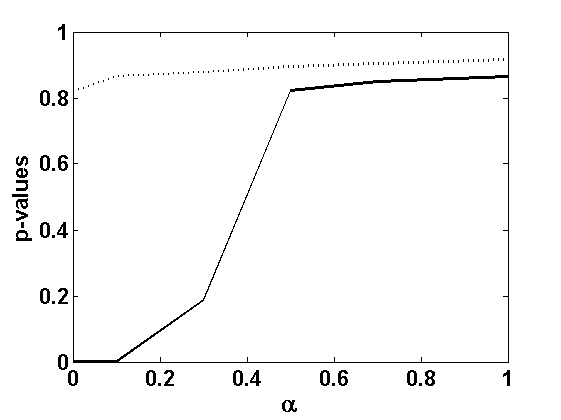}
	}
	\subfloat[Variable: BMI; ~~$H_0 : \sigma=4$]{
		\includegraphics[width=0.3\textwidth]{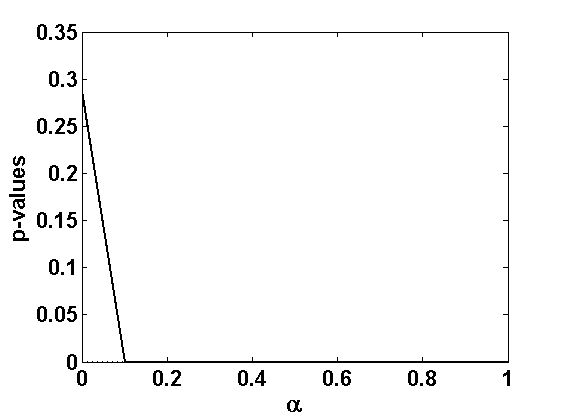}
	}
	\subfloat[Variable: Wt; ~~$H_0 : \mu=72$]{
		\includegraphics[width=0.3\textwidth]{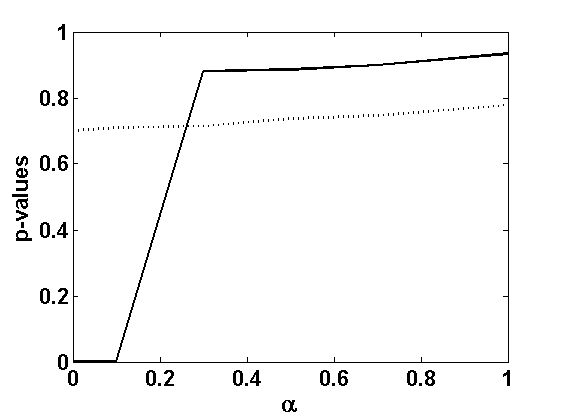}
	}	
	\caption{P-values obtained for different hypotheses testing problems for AIS data using the MDPDE based Wald-type tests 
		for the full data (solid line) and the outlier-deleted data (dotted line).}
	\label{FIG:Data_pValue}
\end{figure}

\begin{center}
	\begin{tabular}{l|llllll}\hline
		Variable & HC & WCC & LBM & PFC & BMI & Wt \\\hline 
		Hypothesis $H_0$  & $\gamma=-1.8$~~ & $\gamma=1.7$~~ & $\gamma=2$~~
		& $ \sigma=57$~~  & $ \sigma=4$~~ & $ \mu=72$\\
		\hline
	\end{tabular}
\end{center}%
%
For all these hypotheses, we have computed the p-values using the MDPDE based Wald-type tests for different $\alpha$
for the full data as well as the outlier-deleted data, which are plotted in Figure \ref{FIG:Data_pValue}.
Note that, the usual Wald test at $\alpha=0$ is strongly affected by the outliers and provides completely opposite inference
with clear difference in significance levels in presence or absence of outliers in most cases. However, the proposed MDPDE based tests with $\alpha>0$ 
provides stable inference similar to the one we could have obtained after removing the outliers 
for all the variables except PFC; for the testing problem in PFC, we need $\alpha\geq 0.2$ to have robust inference
due to the excessive amount of contamination. Another interesting case is the one with LBM,
where the p-values obtained under the full data and outlier removed data are almost the same, 
except for the classical Wald test at $\alpha=0$; 
the corresponding p-values obtained by Wald test are 0.000066 and 0.065, respectively.
Thus, here also the inference at 95\% significance level alters due to the outlier for Wald tests,
but the proposed MDPDE based tests at $\alpha>0$ yield more reasonable inference of failing to reject the hypothesis even in the presence of outliers.
These observations further support our claimed robustness of the proposed MDPDE based Wald-type tests.

\subsection{AIDS Clinical Trial Data}

Our second example is an AIDS clinical trial (ACTG 315) including 46 HIV-1 infected patients treated with 
a potent antiretroviral drug cocktail based on protease inhibitor and reverse transcriptase drugs (ritonavir, 3TC and AZT).
During the study, the viral load, cd4 count (CD4) and cd8 count (CD8) were measured several times 
in different days from the start of the treatment (generally 4 to 10 measurements per patient) 
The corresponding data has been analyzed by several statisticians \cite{Wu/etc:1999, Wu:2002, Lachos/etc:2013}
and is available in the R package '\textit{qrNLMM}'. In particular \citet{Castro/etc:2019}
fitted the skew-normal distribution to this data in a regression settings. 

\begin{figure}[h]
	\centering
	\subfloat[CD4 ]{
		\begin{tabular}{c}			
			\includegraphics[width=0.3\textwidth]{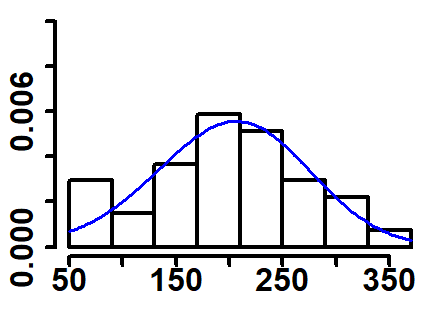}
			\\	\includegraphics[width=0.28\textwidth]{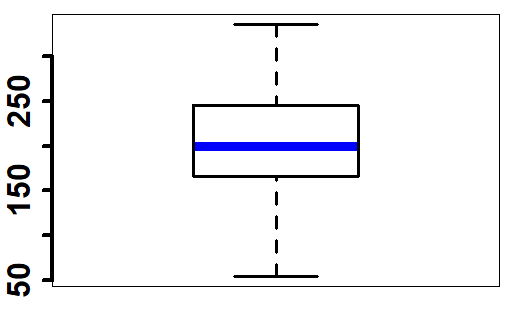}
			\label{FIG:Data_CD4_day_2}
		\end{tabular}
	}
	\subfloat[CD8]{
		\begin{tabular}{c}			
			\includegraphics[width=0.3\textwidth]{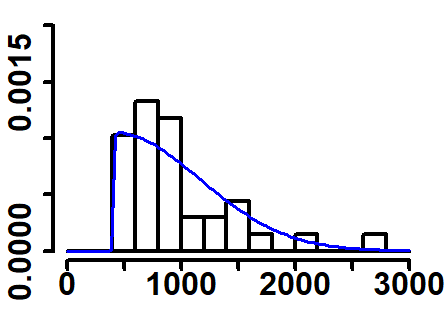}
			\\	\includegraphics[width=0.28\textwidth]{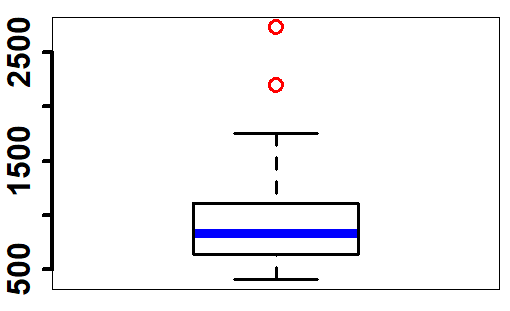}
			\label{FIG:Data_CD8_day_2}
		\end{tabular}
	}
\subfloat[LGVIRAL]{
	\begin{tabular}{c}			
		\includegraphics[width=0.3\textwidth]{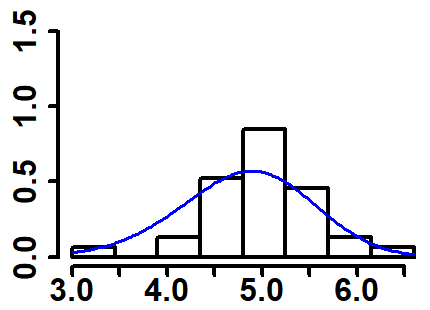}
		\\	\includegraphics[width=0.28\textwidth]{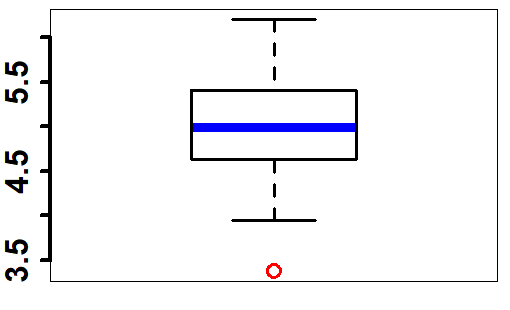}
		\label{FIG:Data_lgviral_day_2}
	\end{tabular}
}
	\caption{Histograms and Box-plots of different variables, measured at day 2, from the ACTG 315 Data.
		The SN distributions fitted by the MLE are also shown along with the histograms.}
	\label{FIG:HIV_Data_hist}
\end{figure}

Here, we consider the variable CD4, CD8 and the logarithm of the viral load (LGVIRAL) measured at 
the second day after the start of the study for each patients. 
The corresponding histogram and the SN fit by the MLE is presented in Figure \ref{FIG:HIV_Data_hist};
clearly the distributions are skewed but the MLE is unable to fit them properly for CD8 and LGVIRAL
due to the presence of outliers as shown in the respective box-plots in the same figure (Figure \ref{FIG:HIV_Data_hist}). 
The MLE based fits are also shown in the figures along with the histogram, which clearly show the inability of the MLE to 
adequately model the bulk of the data due to the presence of few outlying points. 
In particular, the fitted SN distributions (by MLE) have a clearly different mode for both the measurements
CD8 and LGVIRAL due to strong outlier effects.

\begin{figure}[h!]
	\centering
	\subfloat[CD8]{
		\begin{tabular}{c}			
			\includegraphics[width=0.45\textwidth]{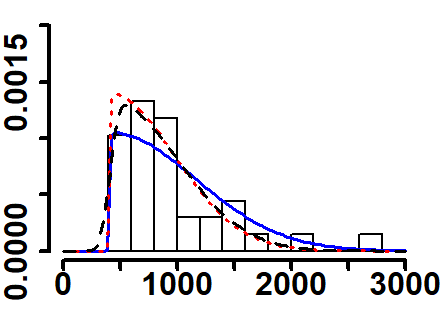}
			\label{FIG:Data_CD8_day_2_dpd}
		\end{tabular}
	}
\subfloat[LGVIRAL]{
	\begin{tabular}{c}			
		\includegraphics[width=0.45\textwidth]{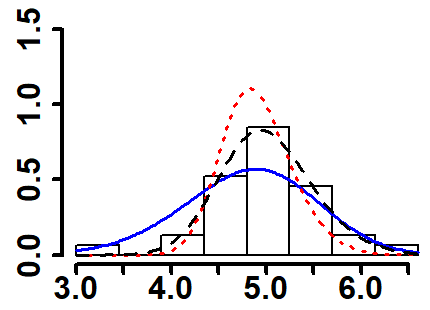}
		\label{FIG:Data_lgviral_day_2_dpd}
	\end{tabular}
}
	\caption{The SN distributions fitted by the MDPDE at $\alpha=0.5$ (black dashed line), 
		the MLE (blue solid line) and the outlier deleted MLE (red dotted line), 
		along with the histograms of the measurements LGVIRAL and CD8 in AIDS clinical trial data.}
	\label{FIG:HIV_Data_hist_dpd}
\end{figure}

We next compute the proposed MDPDEs of the parameters of the fitted SN distribution 
for each of the three health measurements and compared them with the MLEs and the outlier deleted MLEs;
the resulting estimates and their estimated standard errors are presented in Table \ref{TAB:CD4_Data_MDPDE}.
From the table, we can see that the MDPDEs at any $\alpha>0$ are very similar to the MLE for CD4
where there are no outliers in the data. For LGVIRAl, the MDPDEs with $\alpha\geq 0.3$ produce robust results
which are significantly different from the MLE and are close to the outlier deleted MLE.
For CD8, however, the MDPDEs with $\alpha\geq 0.1$ are all similar but significantly different 
from both the MLE as well as the outlier deleted MLE. To see which one provides the more robust fit, 
in Figure \ref{FIG:HIV_Data_hist_dpd}, we have plotted the fitted SN density obtained by 
the MDPDE at $\alpha=0.5$, the MLE and the outlier deleted MLE, along with the histograms of CD8 and LGVIRAL. 
In both cases, the MDPDE seems to provide the best fit to the major bulk of the histogram, 
even better than the outlier deleted MLE. This shows that there are yet other masked outliers in the data
which are not detectable by the usual box-plot technique
and hence illustrates the significance of our proposed MDPDEs over outlier deletion methods
in providing stable inference from contaminated data.

\bigskip
\begin{table}[!h]
	\caption{The parameter estimates (standard errors) for the fitted SN distribution 
		to the three health measurements in AIDS clinical trail data (measured at day 2), 
		obtained through MDPDEs at different $\alpha$, the MLE (at $\alpha=0$) and the outlier deleted MLE.
		The number of outliers found for each measurements are reported after their name in the first column.}
	\centering	
	\resizebox{0.8\textwidth}{!}{
		\begin{tabular}{l c|c c c c c c| c}
			\hline
			Variable &  & & & & $\alpha$ & & & Outlier deleted \\
			(Outlier)&  & 0(MLE) & 0.1 & 0.3 & 0.5 & 0.7 & 1.0 & MLE\\
			\hline
		CD4  & $\mu$ & 252.650	 & 252.634	& 252.634	& 252.634	& 252.634	& 252.634	& 252.650\\
			(0) & & (1.538)	& (1.729)	& (1.954)	& (2.177) & (2.280)	& (2.470)	& (1.538)\\
			& $\sigma$	& 89.204 & 89.204 & 89.204 & 89.204 & 89.204	& 89.204 & 89.204\\
			& & (1.169) & (1.243) & (1.290) & (1.343) & (1.475) & (1.989) & (1.160)\\
			& $\gamma$ & -1.084	 & -1.057 & -1.058 & -1.059 & -1.059 & -1.059 & -1.084\\
			& & (0.675) & (0.696) & (0.733) & (0.738) & (0.752) & (0.755) & (0.679)\\
			\hline				
			CD8	&	$\mu$	& 407.251	&	407.141	&	407.140	&	407.140	&	407.140	&	407.140	&	408.972	\\
(2)	&	&	(1.605)	&	(1.736)	&	(1.804)	&	(2.063)	&	(2.160)	&	(2.708)	&	(1.284)	\\
&	$\sigma$ & 757.601	&	594.563	&	594.562	&	594.562	&	594.562	&	594.562	&	570.614	\\
&	&	(1.839)	&	(2.297)	&	(2.406)	&	(2.749)	&	(3.386)	&	(3.986)	&	(1.771)	\\
&	$\gamma$ & 109.581	&	9.739	&	9.738	&	9.738	&	9.738	&	9.738	&	87.762	\\
&	&	(1.291)	&	(2.264)	&	(2.370)	&	(2.442)	&	(2.751)	&	(3.216)	&	(0.790)	\\
\hline
LGVIRAL	&	$\mu$ & 5.374	&	4.548	&	4.592	&	4.598	&	4.597	&	4.595	&	4.558	\\
(1)	&	&	(0.076)	& (0.076)	&	(0.078)	&	(0.086)	&	(0.105)	&	(0.278)	&	(0.053)	\\
&	$\sigma$ & 0.902	&	0.713	&	0.675	&	0.642	&	0.630	&	0.620	&	0.510	\\
&	&	(0.076)	&	(0.079)	&	(0.085)	&	(0.096)	&	(0.117)	&	(0.166)	&	(0.055)	\\
&	$\gamma$ & -1.231	&	1.222	&	1.271	&	1.380	&	1.485	&	1.609	&	1.679	\\
&	&	(0.160)	&	(0.305)	&	(0.326)	&	(0.330)	&	(0.331)	&	(0.927)	&	(0.344)	\\
\hline																
	\end{tabular}}
	\label{TAB:CD4_Data_MDPDE}
\end{table}

Finally, as in the previous example, here also we have observed that 
the MDPDEs with $\alpha\geq 0.3$ are extremely stable in the presence and absence of the outliers, 
and produce robust inference for any parametric hypothesis testing problem for these clinical trial data as well.
So, we have not presented them here for brevity.

\section{Concluding remarks}

In this paper, we have discussed new robust inference procedures for the SN distribution
which is useful in modeling noisy skewed data through the popular minimum DPD approach.
The minimum DPD estimators of the SN parameter are described along with their asymptotic and robustness properties
and two efficient computational algorithms have been proposed. 
Then, we discuss the robust testing procedure through the MDPDE based Wald-type tests and their properties
with detailed illustrations for testing symmetry against SN alternatives.
The usefulness of the SN distribution and the proposed robust inference 
in the context of health data analysis are argued and illustrated empirically.

This work opens up several new directions in health research.
The proposed methodology is a generalization of the MLE
which can be extended to different inferential problems in health studies with skewed data to generate stable insight.
For example, the immediate extension would be robust inference under regression models for skewedly distributed responses
or comparing different populations of skewed data. 
The latter can be used in finding differential genes from expression data which are skewed in nature.

Although we have suggested some empirical choices for the tuning parameter $\alpha$ to be used in practice,
more detailed research in this line would be necessary to develop an algorithm for data-driven selection of $\alpha$;
see, for example, \cite{Warwick/Jones:2005}.
We hope to pursue some of these extensions in our future work.

\bigskip\noindent\textbf{Acknowledgment:}\\
The research of the third author (AG) is partially supported by the INSPIRE Faculty Research Grant
from  Department of Science and Technology, Government of India,
and the research of the first and third authors (AN and AG) are partially 
supported by a Start-up Research Grant from Indian Statistical Institute.


\begin{thebibliography}{}
	%
	%
	
\bibitem[\protect\citeauthoryear{Azzalini}{Azzalini}{1985}]{Azzalini:1985}
Azzalini A. 
A class of distributions which includes the normal ones. 
\textit{Scand J Stat} 1985; 12:171--178.

\bibitem[\protect\citeauthoryear{Azzalini}{Azzalini}{1986}]{Azzalini:1986}
Azzalini A.
Further results on a class of distributions which includes the normal ones. 
\textit{Statistica} 1986; 46:199--208.

\bibitem[\protect\citeauthoryear{Azzalini}{Azzalini}{2005}]{Azzalini:2005}
 Azzalini A.
 The skew-normal distribution and related multivariate families. 
 \textit{Scand J Stat} 2005; 32:159--188.

\bibitem[\protect\citeauthoryear{Azzalini}{Azzalini}{2011}]{Azzalini:2011}
Azzalini A. 
Skew-symmetric families of distributions. 
\textit{Int Encyclopedia Statist Sci} 2011; 1344--1346.

\bibitem[\protect\citeauthoryear{Hossain,A. and Beyene,}{Hossain and Beyene}{2015}]{Hossain/Beyene:2015}
Hossain A and Beyene J. 
Application of Skew Normal Distribution for Detecting Differential Expression to micro-RNA Data. 
\textit{J Appl Stat}, 2015; 42(3):477--491


\bibitem[\protect\citeauthoryear{Castro}{Castro}{2019}]{Castro/etc:2019}
Castro LM, Wang WL, Lachos VH, Inacio de Carvalho V and Bayes CL.  
Bayesian semiparametric modeling for HIV longitudinal data with censoring and skewness. 
{\it Statist Methods Med Res} 2019; 28(5):1457-1476.	


\bibitem[\protect\citeauthoryear{Warwick and Jones}{Warwick and Jones}{2005}]{Ferreira/etc:2018}
da Silva Ferreira C, Vilca F and Bolfarine H.  
Diagnostics analysis for skew-normal linear regression models: Applications to a quality of life dataset. 
{\it Brazilian J Prob Stat} 2018; 32(3):525-544.	


\bibitem[\protect\citeauthoryear{Ghalani, M. R., and Zadkarami}{Ghalani and Zadkarami}{2019}]{Ghalani/Zadkarami:2019}
Ghalani MR and Zadkarami MR.  
Investigation of covariance structures in modelling longitudinal ordinal responses with skew normal random effect. 
{\it Communic Stat - Simul Comput}, 2019; 1-16.


\bibitem[\protect\citeauthoryear{Yiu, S., and Tom}{Yiu, S., and Tom}{2018}]{Yiu/Tom:2018}
Yiu S and Tom BD.  
Two-part models with stochastic processes for modelling longitudinal semicontinuous data: Computationally efficient inference and modelling the overall marginal mean. 
{\it Statist Methods Med Res} 2018; 27(12):3679-3695.


\bibitem[\protect\citeauthoryear{Liu}{Liu et al.}{2016}]{Liu/etc:2016}
Liu L, Strawderman RL, Johnson BA  and  O'Quigley JM.  
Analyzing repeated measures semi-continuous data, with application to an alcohol dependence study. 
{\it Statist Methods Med Res} 2016; 25(1):133-152.


\bibitem[\protect\citeauthoryear{Smith}{Smith et al.}{2017}]{Smith/etc:2017}
Smith VA, Neelon B, Preisser JS and Maciejewski ML.  
A marginalized two-part model for longitudinal semicontinuous data. 
{\em Statist Methods Med Res} 2017; 26(4):1949-1968.


\bibitem[\protect\citeauthoryear{Wason, J. M., and Mander}{Wason and Mander}{2015}]{Wason/Mander:2015}
Wason JM and Mander AP. 
The choice of test in phase II cancer trials assessing continuous tumour shrinkage when complete responses are expected. 
{\it Statist Methods Med Res} 2015; 24(6):909-919.


\bibitem[\protect\citeauthoryear{Gutman, R., and Rubin}{Gutman, R., and Rubin}{2017}]{Gutman/Rubin:2017}
Gutman R and Rubin DB.
Estimation of causal effects of binary treatments in unconfounded studies with one continuous covariate. 
{\it Statist Methods Med Res} 2017; 26(3):1199-1215.


\bibitem[\protect\citeauthoryear{Xing}{Xing et al.}{2017}]{Xing/etc:2017}
Xing D, Huang Y, Chen H, Zhu Y, Dagne GA and Baldwin J.  
Bayesian inference for two-part mixed-effects model using skew distributions, with application to longitudinal semicontinuous alcohol data. 
{\it Statist Methods Med Res} 2017; 26(4):1838-1853.

%

\bibitem[\protect\citeauthoryear{Sengupta}{Sengupta et al.}{2015}]{Sengupta/etc:2015}
Sengupta D, Choudhary PK and Cassey P. 
Modeling and Analysis of Method Comparison Data with Skewness and Heavy Tails. 
In {\it Ordered Data Analysis, Modeling and Health Research Methods} 2015; (pp. 169-187). Springer, Cham.


\bibitem[\protect\citeauthoryear{Crocetta, C., and Loperfido}{Crocetta and Loperfido}{2009}]{Crocetta/Loperfido:2009}
Crocetta C and Loperfido N.  
Maximum likelihood estimation of correlation between maximal oxygen consumption and the 6-min walk test in patients with chronic heart failure. 
{\it J Appl Stat} 2009; 36(10):1101-1108.


\bibitem[\protect\citeauthoryear{van den Hout, A., and Matthews}{van den Hout and Matthews}{2009}]{vandenHout/Matthews:2009}
van den Hout A and Matthews FE.  
A piecewise-constant Markov model and the effects of study design on the estimation of life expectancies in health and ill health. 
{\it Statist Methods Med Res} 2009; 18(2):145-162.


\bibitem[\protect\citeauthoryear{Bandyopadhyay}{Bandyopadhyay et al.}{2010}]{Bandyopadhyay/etc:2010}
Bandyopadhyay D, Lachos VH, Abanto‐Valle CA and Ghosh P.  
Linear mixed models for skew‐normal/independent bivariate responses with an application to periodontal disease. 
{\it Stat Med} 2010; 29(25):2643-2655.


\bibitem[\protect\citeauthoryear{Smirnova}{Smirnova et al.}{2018}]{Smirnova/etc:2018}
Smirnova E, Huzurbazar S and Jafari F. PERFect: PERmutation Filtering test for microbiome data. 
\textit{Biostatistics}. 2018 Jun 18.


\bibitem[\protect\citeauthoryear{Daly}{Daly et al.}{2017}]{Daly/etc:2017}
Daly CH, Higgins V, Adeli K, Grey VL and Hamid JS. 
Reference interval estimation: methodological comparison using extensive simulations and empirical data. 
\textit{Clin Biochem} 2017; 50(18):1145--1158.


\bibitem[\protect\citeauthoryear{Ngunkeng}{Ngunkeng}{2013}]{Ngunkeng:2013}
Ngunkeng G. 
Statistical analysis of skew normal distribution and its applications.
Doctoral dissertation, Bowling Green State University; 2013.


\bibitem[\protect\citeauthoryear{Giuntella}{Giuntella}{2017}]{Giuntella:2013}
Giuntella O. 
Why does the health of immigrants deteriorate? Evidence from birth records.
\textit{J Health Econ.} 2017; 54:1--16.

\bibitem[\protect\citeauthoryear{Huang, C. Y., and Ku}{Huang and Ku}{2010}]{Huang/Ku:2010}
Huang CY and Ku MS. 
Asymmetry effect of particle size distribution on content uniformity and over-potency risk in low-dose solid drugs. 
{\em J Pharm Sci} 2010; 99(10):4351--4362.

\bibitem[\protect\citeauthoryear{Telford, R.D., Cunningham}{Telford and Cunningham}{1991}]{Telford/Cunningham:1991}
Telford RD and  Cunningham RB. 
Sex, Sport, and Body-size Dependency of Hematology in Highly Trained Athletes. 
\textit{Med Sci Sports Exerc} 1991; 23(7):788--794


\bibitem[\protect\citeauthoryear{Yal\c{c}inkaya..}{Yal\c{c}inkaya et al.}{2017}]{Yalcinkaya/etc:2017}
Yal\c{c}inkaya A, Şenoglu B, Yolcu U. 
Maximum likelihood estimation for the parameters of skew normal distribution using genetic algorithm. 
\textit{Swarm  Evolut Comput}. 2018; 38:127--38.	



\bibitem[\protect\citeauthoryear{Azzalini and Regoli}{Azzalini and Regoli}{2012}]{Azzalini/Regoli:2012}
Azzalini A and Regoli G. 
The work of Fernando de Helguero on non-normality arising from selection. 
\textit{Chilean J Stat} 2012; 3(2).


\bibitem[\protect\citeauthoryear{Basso}{Basso et al.}{2010}]{Basso/etc:2010}
Basso RM, Lachos VH, Cabral CRB and Ghosh P.  
Robust mixture modeling based on scale mixtures of skew-normal distributions. 
{\it Comput Stat Data Anal} 2010; 54(12):2926--2941.


\bibitem[\protect\citeauthoryear{Zeller}{Zeller et al.}{2016}]{Zeller/etc:2016}
Zeller CB, Cabral CR and Lachos VH. 
Robust mixture regression modeling based on scale mixtures of skew-normal distributions. 
{\it TEST} 2016; 25(2):375-396.

\bibitem[\protect\citeauthoryear{Hashimoto}{Hashimoto}{2017}]{Hashimoto:2017}
Hashimoto S.  
Robust estimation of skew-normal distribution with location and scale parameters via log-regularly varying functions. 
{\it Int J Stat Syst} 2017; 12(4):813-822.

\bibitem[\protect\citeauthoryear{Nurminen}{Nurminen et al.}{2015}]{Nurminen/etc:2015}
Nurminen H, Ardeshiri T, Piche R and Gustafsson F.  
Robust inference for state-space models with skewed measurement noise. 
{\it IEEE Signal Proces Lett} 2015; 22(11):1898--1902.


\bibitem[\protect\citeauthoryear{Basu, Harris, Hjort and Jones}{Basu et~al.}{1998}]{Basu/etc:1998}
Basu A, Harris IR, Hjort NL, Jones MC. 
Robust and efficient estimation by minimising a density power divergence. 
\textit{Biometrika} 1998; 85(3):549--59.


\bibitem[\protect\citeauthoryear{Basu, Shioya and Park}{Basu et~al.}{2011}]{Basu/etc:2011}
Basu A, Shioya H, Park C. 
\textit{Statistical inference: the minimum distance approach. }
Chapman and Hall/CRC; 2011.


\bibitem[\protect\citeauthoryear{Basu et al}{Basu et~al.}{2016}]{Basu/etc:2016}
Basu A, Mandal A, Martin N, Pardo L. 
Generalized Wald-type tests based on minimum density power divergence estimators. 
\textit{Statistics} 2016; 50(1):1--26.

\bibitem[\protect\citeauthoryear{Ghosh and Basu}{Ghosh and Basu}{2013}]{Ghosh/Basu:2013}
Ghosh A, Basu A. 
Robust estimation for independent non-homogeneous observations using density power divergence with applications to linear regression. 
\textit{Electron J Stat} 2013; 7:2420--56.

\bibitem[\protect\citeauthoryear{Ghosh and Basu}{Ghosh and Basu}{2015}]{Ghosh/Basu:2015}
Ghosh A and Basu A. 
Robust Estimation for Non-Homogeneous Data and the Selection of the Optimal Tuning Parameter: The DPD Approach. 
\textit{J Appl Stat} 2015; 42(9):2056---2072. 


\bibitem[\protect\citeauthoryear{Ghosh and Basu}{Ghosh and Basu}{2017}]{Ghosh/Basu:2017}
Ghosh A and Basu A. 
Robust and efficient parameter estimation based on censored data with stochastic covariates.
{\it Statistics} 2017; 51(4):801--823..

\bibitem[\protect\citeauthoryear{Ghosh and Basu}{Ghosh et al.}{2019}]{Ghosh/etc:2017}
Ghosh A, Basu A and Pardo L. 
Robust Wald-Type Tests under Random Censoring with Applications to Clinical Trial Analyses.
{\it ArXiv Pre-print}, 2019; arXiv:1708.09695v2 [stat.ME]. 


\bibitem[\protect\citeauthoryear{Ghosh}{Ghosh}{2019}]{Ghosh:2019}
Ghosh A.  
Robust inference under the beta regression model with application to health care studies. 
{\em Statist Methods Medical Res} 2019; 28(3):871-888.

\bibitem[\protect\citeauthoryear{Ghosh and Basu}{Ghosh et al.}{2016}]{Ghosh/etc:2016}
Ghosh A, Mandal A, Martin N and Pardo L.
Influence Analysis of Robust Wald-type Test.
\textit{J Mult Anal} 2016; 147:102--126


\bibitem[\protect\citeauthoryear{Sivananadam}{Sivananadam and Deepa}{2008}]{Sivananadam/Deepa:2008}
Sivananadam SN and Deepa SN
\textit{Introduction to Genetic Algorithm}, Springer-Verlag Berlin Heidelberg; 2008.




\bibitem[\protect\citeauthoryear{Hampel, Ronchetti, Rousseeuw, and Stahel}{Hampel et~al.}{1986}]{Hampel/etc:1986}
Hampel FR, Ronchetti E, Rousseeuw PJ, et al. 
\textit{Robust Statistics: The Approach Based on Influence Functions}. 
New York, USA: John Wiley \& Sons, 1986. 


\bibitem[\protect\citeauthoryear{Goldberg}{Goldberg}{1989}]{Goldberg:1980}
Goldberg  DE. 
\textit{Genetic Algorithm in Search, Optimization and Machine Learning}, 
Addison-Wesley Publishing Company Inc.; 1989.



\bibitem[\protect\citeauthoryear{Chudasama  ..}{Chudasama et al.}{2011}]{Chudasama/etc:2011}
Chudasama C, Shah SM, Panchal M. Comparison of parents selection methods of genetic algorithm for TSP. 
In \textit{International Conference on Computer Communication and Networks} CSI-COMNET-2011, Proceedings 2011 (pp. 85-87).


\bibitem[\protect\citeauthoryear{Kim}{Kim et. al.}{2016}]{Kim/etc:2016}
Kim D and Fessler JA.
Optimized first-order methods for smooth convex minimization.
{\it Math Prog} 2016; 151(1-2):81--107.

\bibitem[\protect\citeauthoryear{Snyman}{Snyman et. al.}]{Snyman/etc:2018}
Snyman JA and Wilke DN.
Practical Mathematical Optimization- Basic Optimization Theory and Gradient based Algorithms.
{\it Springer Optim Appl} 2018; 133(2 ed.), Springer.


\bibitem[\protect\citeauthoryear{Vandenberghe}{Vandenberghe}{2019}]{Vandenberghe/etc:2019} 
Vandenberghe L.
Fast Gradient Methods.
{\it Lecture notes for EE236C at UCLA}; 2019.

\bibitem[\protect\citeauthoryear{Robinns}{Robinns et.al.}]{Robinns/etc:1951}
Robins H and Monro S.
A stochastic approximation method.
{\it Annal Math Stat} 1951; 22:400--407.

\bibitem[\protect\citeauthoryear{Barzilai}{Barzilai et al.}{1988}]{Barzilai/etc:1988}
Barzilai J and Borwein JM.
Two-point step size Gradient Methods.
{\it IMA J Numer Anal} 1988; 8(1):141--148.

\bibitem[\protect\citeauthoryear{Yuan}{Yuan}]{Yuan/etc:1999}
Yuan Y.
Step-sizes for the Gradient Method.
{\it AMS/IP Studies Adv Math} 1999; 42(2):785--805



\bibitem[\protect\citeauthoryear{Wu}{Wu}{2002}]{Wu:2002}
Wu L.  
A joint model for nonlinear mixed-effects models with censoring and covariates
measured with error, with application to aids studies. 
{\it J Amer Statist Assoc} 2002; 97(460):955--964.

\bibitem[\protect\citeauthoryear{Lachos}{Lachos et al.}{2013}]{Lachos/etc:2013}
Lachos VH, Castro LM and Dey DK.  
Bayesian inference in nonlinear mixed-effects models using normal independent distributions. 
{\it Comput Stat Data Anal} 2013; 64:237--252.


\bibitem[\protect\citeauthoryear{Wu}{Wu}{1999}]{Wu/etc:1999}
Wu H and Ding AA.
Population HIV-1 Dynamics In Vivo: Applicable Models and Inferential Tools for Virological Data from AIDS Clinical Trials.
{\it Biometrics} 1999; 55(2):410--418.

	
\bibitem[\protect\citeauthoryear{Warwick and Jones}{Warwick and Jones}{2005}]{Warwick/Jones:2005}
Warwick J and Jones MC. 
Choosing a robustness tuning parameter.         
\textit{J Stat Comput Simul} 2005; 75:581--588.




	


	



%










\end{thebibliography}
\end{document}